\documentclass[11pt]{article}      
\pdfoutput=1
\usepackage{amsthm,amsmath,amssymb,xcolor,sectsty,latexsym,mathtools,hyperref,mathrsfs}
\usepackage[shortlabels]{enumitem}
\usepackage{slashed}  
\usepackage{bm}
\usepackage{epsfig}
\usepackage{dcolumn}
\definecolor{vermelho}{cmyk}{0,.88,.77,.40}
\usepackage{cite}
\usepackage{feynmp}
\numberwithin{equation}{section}
\usepackage[textwidth=6.5in,top=1.2in,bottom=1.2in]{geometry}
\usepackage{setspace}
\chapterfont{\color{vermelho}}
\usepackage{simplewick}
\usepackage{hyperref}
\usepackage{float}
\usepackage{mathtools}
\usepackage{blkarray, bigstrut} %

\newcommand{\be}{\begin{equation}}
\newcommand{\ee}{\end{equation}}
\newcommand{\beq}{\begin{equation}}
\newcommand{\eeq}{\end{equation}}
\newcommand{\ba}{\begin{eqnarray}}
\newcommand{\ea}{\end{eqnarray}}
\newcommand{\bef}{\begin{figure}}
\newcommand{\eef}{\end{figure}}
\newcommand{\amp}{&\!\!\!}

\newcommand{\p}{\partial}

\newcommand{\cO}{{\cal O}}

\newcommand{\cL}{{\cal L}}

\newcommand{\cH}{{\cal H}}

\def\q{{\bf q}}
\def\x{{\bf x}}
\def\y{{\bf \hat{x}}}
\def\k{{\bf k}}

\def\pv{{\bf p}}
\def\rv{{\bf r}}
\def\bv{{\bf b}}
\def\Lv{{\bf L}}

\newcommand{\bra}[1]{\left\langle #1 \right|}
\newcommand{\ket}[1]{\left|#1\right\rangle}
\newcommand{\braket}[2]{\left\langle {#1} \middle| {#2} \right\rangle}

\newcommand{\mpl}{M_{Pl}}

\newcommand{\dv}{\hat{p}}
\newcommand{\DM}{\text{DM}}
\newcommand{\ch}{\chi}

\newcommand{\Gamd}{\Gamma_{\mbox{\tiny{dec}}}}
\newcommand{\tid}{t_{\mbox{\tiny{dec}}}}

\newcommand{\qt}{\tilde{q}}
\newcommand{\qtv}{{\bf\tilde{q}}}
\newcommand{\sigmat}{\tilde{\sigma}}

\newcommand{\oma}{\omega}
\newcommand{\ma}{m_a}
\newcommand{\mpr}{m_p}
\newcommand{\scal}{\chi}
\newcommand{\rg}{\sqrt{-g}}
\newcommand{\F}{F}

\newcommand{\Ek}{{E_k}}
\newcommand{\Eq}{{E_q}}
\newcommand{\psik}{{\tilde{\psi}_\k}}
\newcommand{\zeto}{\xi}
\newcommand{\press}{P}
\newcommand{\A}{V^{(0)}}
\newcommand{\B}{V^{(2)}}
\newcommand{\AB}{V^{(n)}}
\newcommand{\Ahat}{\hat{V}^{(0)}}
\newcommand{\Bhat}{\hat{V}^{(2)}}
\newcommand{\rhoA}{\rho^{(0)}}
\newcommand{\rhoB}{\rho^{(2)}}
\newcommand{\rhoAB}{\rho^{(n)}}
\newcommand{\PsiA}{\Psi^{(0)}}
\newcommand{\PsiB}{\Psi^{(2)}}
\newcommand{\PsiAB}{\Psi^{(n)}}
\newcommand{\deltaA}{\delta^{(0)}}
\newcommand{\deltaB}{\delta^{(2)}}
\newcommand{\deltaAB}{\delta^{(n)}}

\newcommand{\scprof}{\phi_s}

\newcommand{\bt}{\tilde{\beta}}

\newcommand{\en}{u}
\newcommand{\phia}{\phi_a}
\newcommand{\abc}{K}
\newcommand{\an}{a}
\newcommand{\bn}{b}
\newcommand{\cn}{c}
\newcommand{\phis}{\Theta}
\newcommand{\sig}{\bar{\sigma}}

\begin{document}

\thispagestyle{empty}
\begin{titlepage}
\nopagebreak

\title{  \begin{center}\bf Decoherence from General Relativity\end{center} }

\vfill
\author{Itamar J.~Allali\footnote{itamar.allali@tufts.edu}, ~ Mark P.~Hertzberg\footnote{mark.hertzberg@tufts.edu}}
\date{ }

\maketitle

\begin{center}
	\vspace{-0.7cm}
	{\it  Institute of Cosmology, Department of Physics and Astronomy}\\
	{\it  Tufts University, Medford, MA 02155, USA}
	\end{center}
%\vfill
\bigskip

\begin{abstract}
It is of great interest to explore matter in nontrivial quantum arrangements, including Schr\"{o}dinger cat-like states. Such states are sensitive to decoherence from their environment. Recently, in Ref.~\cite{Allali:2020ttz} we computed the rate of decoherence of a piece of superposed matter that primarily only interacts gravitationally, a dark-matter-Schr\"{o}dinger-cat-state (DMSCS), within the nonrelativistic approximation. In this work we improve this to a general relativistic analysis. We firstly derive a single particle relativistic Schr\"{o}dinger equation for a probe particle that passes through the DMSCS; the interaction is provided by the weak field metric of general relativity from the source. For a static DMSCS we find a neat generalization of our previous results. We then turn to the interesting new case of a time dependent DMSCS, which can be provided by a coherently oscillating axion field leading to superposed time dependent oscillations in the metric; a truly quantum-general relativistic phenomenon. We use scattering theory to derive the decoherence rate in all these cases. When the DMSCS is in a superposition of distinct density profiles, we find that the decoherence rate can be appreciable. We then consider the novel special case in which the density is not in a superposition, but the phase of its field oscillation is; this is a property that cannot be decohered within the nonrelativistic framework. We find that if the probe particle and/or the DMSCS's velocity dispersion is slow, then the rate of decoherence of the phase is exponentially suppressed. However, if both the probe and the DMSCS's velocity dispersion are relativistic, then the phase can decohere more rapidly. As applications, we find that diffuse galactic axions with superposed phases are robust against decoherence, while dense boson stars and regions near black hole horizons are not, and we discuss implications for experiment.
\end{abstract}

\end{titlepage}

\setcounter{page}{2}

\tableofcontents

\section{Introduction}
It is interesting to explore novel phenomena in which both gravitation and quantum mechanics play the central role. For most ordinary matter, although their quantum character can be made manifest, other interactions, such as electromagnetic, often play the central role in its dynamics. To isolate the interplay between gravitational and quantum mechanical phenomena, it is useful to consider matter that has primarily only gravitational interactions. This is not known among the familiar particles, but may well be the most dominant form of matter in the universe (for a review, see Ref.~\cite{Peebles:2013hla}). Further, we can simply consider the presence of such exotic matter, even if it were not the dominant component. In any case, we shall refer to such material that primarily interacts via gravitation, as ``dark matter" (DM) in this work.

The current lack of direct detection of the DM that actually does make up most of the mass of the universe, other than its gravitational impact on galaxies, etc., implies that DM is at most very weakly coupled to the ordinary particles of the Standard Model. In fact it is possible that its only coupling to the Standard Model is gravitational (plus other Planck, or nearly Planck, suppressed operators). On the one hand, since the interactions between DM particles and ordinary matter particles is very weak and/or infrequent it makes it very difficult to detect the DM properties. On the other hand, this may offer a new opportunity for rich behavior for the following reason: due to the lack of significant interactions, DM could possibly possess long-lived exotic quantum mechanical phenomena. In particular, one can imagine that a piece of DM has organized into a macroscopic superposition of states, which are sometimes referred to as ``Schr\"{o}dinger cat'' states. 

These Schr\"{o}dinger cat states possess truly quantum behavior as encoded in the off diagonal terms of the density matrix (in the relevant basis). However, the quantumness is usually short-lived for ordinary matter due to interactions with its environment, leading to suppression of the off diagonal elements of the density matrix; a process called \textit{decoherence}. Decoherence can be understood as follows: the Schr\"{o}dinger cat state and its environment interact and inevitably become entangled. The full system remains in a pure state, but an observer will usually not track the full system in all its detail. Instead, one adopts a coarse grained point of view in which the degrees of freedom of the environment (which are typically numerous) are ignored and traced out. This effectively destroys the quantum coherence of the residual sub-system (for early work establishing the mechanism of decoherence, see Refs.~\cite{Zeh1970,Zurek1981,Zurek1982} and for various developments, see Refs.~\cite{Joos:1984uk,Gallis1990,Diosi1995,Giulini:1996nw,Kiefer:1997hv,Dodd:2003zk,Hornberger,Schlosshauer:2003zy,SchlosshauerBook,HornbergerIntrp,Schlosshauer:2019ewh,Nagele:2020kef}). Due to decoherence, the Schr\"{o}dinger cat state evolves into a mixed-state of essentially classical probabilities, rather than a pure quantum mechanical superposition, and so the uniquely quantum mechanical phenomena, such as interference, are no longer present in the reduced system. This process is efficient when the interactions are large, as is the case from an environment of air scattering off some ordinary material. However, this decoherence may be inefficient for DM which lacks these interactions.

There has been much work done on decoherence in the context of gravitation and cosmology; see Refs.~\cite{Bassi:2017szd,Belenchia:2018szb,Asprea:2019dok,Anastopoulos:2013zya,Blencowe:2012mp,Breuer:2008rh,Shariati:2016mty,DeLisle:2019dyw,Orlando:2016pwg,Pang:2016foq,Oniga:2015lro,Bonder:2015hja,Diosi:2015vra,Colin:2014vfa,Hu:2014kia,Pikovski:2013qwa,Polarski:1995jg,Halliwell:1989vw,Kiefer:1998qe,Padmanabhan:1989rm,Kafri:2014zsa,Nelson:2016kjm,Anastopoulos:2014yja,Wang:2006vh,Kok:2003mc,Pikovski:2015wwa,Kiefer:1999gt,Mavromatos:2007hv,Tegmark:2011pi,Anastopoulos:1995ya,Colin:2014cfa,Kiefer:2008ku,Brandenberger:1990bx,Khosla:2016tss,Podolskiy:2015wna,Arrasmith:2017ogi,Albrecht:2018prr}. For DM it is plausible that gravitational or self-interactions can lead to such Schr\"{o}dinger cat-like states due to macroscopic spreading of the wave function, especially if the system exhibits some form of chaos (e.g., see Ref.~\cite{Albrecht:2012zp}). In the case of axion DM \cite{Peccei:1977hh,Weinberg:1977ma,Wilczek:1977pj,Preskill:1982cy,Abbott:1982af,Dine:1982ah,AxionBook,Jaeckel:2010ni}, one is often studying particles of very low mass and so they must have a very high number density to be all the DM. They are therefore normally in very high occupancy states, and so they are usually thought to be classical (e.g., see Refs.~\cite{Davidson:2014hfa,Guth:2014hsa}), but any chaos can lead to the formation of Schr\"{o}dinger cat-like states. So the high occupancy is not a sufficient condition for classicality. Nevertheless, there can be forms of ensemble averaging of classical trajectories that approximately reproduce some quantum correlation functions; see Ref.~\cite{Hertzberg:2016tal}. The residual true quantumness of such states is nontrivial to probe experimentally, but it is conceivable and interesting to consider. So a key issue is whether its quantum character persists or decoheres due to some astrophysical environment.

The rate of decoherence for a {\em dark-matter-Schr\"{o}dinger-cat-state} (DMSCS) of a localized mass distribution for nonrelativistic DM from nonrelativistic probes was computed by us for the first time in Ref.~\cite{Allali:2020ttz}. The superposition involved two differing mass distributions, and the interaction between the environment and the DM was modeled by Newtonian gravity. It was found that such a gravitational interaction would in fact lead to decoherence of the DMSCS depending on parameters. For light bosonic models for DM, such as the axion, the characteristic decoherence time-scales were found to be very sensitive to the mass of the DM particle, with lighter particles leading to rapid decoherence and heavier particles leading to very slow decoherence. The full details were provided by us in Ref.\cite{Allali:2020ttz}, including the dependence on the mass distribution, a comparison between DM in the halo versus near the earth, and so on. Furthermore, when alllowing for spreading of the DMSCS, it was found that most configurations would decohere in times shorter than the age of the universe, except for heavier axions.

One may wonder if general relativistic effects can lead to new interesting forms of decoherence. It is anticipated that any DMSCS which undergoes decoherence within the Newtonian approximation will do the same if one includes relativistic corrections, even if the corrections are appreciable; this is because decoherence is a very robust phenomenon. However, one may consider a DMSCS in which the Newtonian treatment is unable to probe decoherence. In particular, consider a superposition of two states with identical (or nearly identical) Newtonian interactions, but which have distinct behavior within general relativity. A concrete example is a DM source made out of a coherently oscillating scalar field (like an axion) that is in a superposition of different phases of oscillation, but otherwise has the same spatial profile. In this case, the Newtonian interaction would not lead to decoherence, because the Newtonian treatment does not differentiate between the sub-states of the superposition, while the general relativistic treatment does. 

In this work, we consider the mechanism of decoherence of a DMSCS within a fully relativistic treatment, working to linear order in the metric perturbations. We develop a broad formalism for computing the rate of decoherence in this relativistic setup. We first treat the case when the DMSCS generates a static spacetime metric, finding many similarities to the Newtonian formalism in Ref.~\cite{Allali:2020ttz}, but with all the relativistic corrections included. We then generalize this to the case of a time-varying metric, motivated by the light bosonic DM models (e.g., axions), which typically take the form of a regularly oscillating field which source a time-varying metric. The analysis of the time-varying oscillating source is broken up into two cases: the first where the light bosonic DM field has somewhat distinct spatial dependence between the two superposition states, and the second where the superposition sub-states only differ in the phases of the field. This second case captures the main goal of this paper, which is to analyze a situation in which the Newtonian gravitational interaction is completely blind to the superposition, requiring the general relativistic analysis. Within this second case, we provide a characteristic example of a specific configuration of the axion field and discuss the implications of such a DMSCS. We apply these results in several situations, including slowly moving DM in the galaxy and more exotic applications of DM which is relativistic in special environments, such as the near-horizon region of black holes or DM which forms dense boson stars, and we comment on consequences for experiment.

Our paper is organized as follows: in Section~\ref{sec:relschro}, we derive the relativistic Schr\"{o}dinger equation necessary to track the evolution of the environment particles in the spacetime generated by the DM. 
In Section~\ref{ScatDecGeneral} we lay out the basic scattering theory and relation to decoherence needed for later sections.
In Section~\ref{sec:static}, we analyze the evolution of the environment particles in a static spacetime background and apply this analysis to compute the decoherence rate for the DMSCS. In Section~\ref{sec:timedep}, we provide a similar analysis for the case of a time-varying metric, compute the decoherence rate, and also apply these results to a specific profile. In Section~\ref{sec:disc} we apply our results to DM in the galaxy today.
In Section~\ref{sec:outlook} we discuss the spreading of states, application to boson stars, black holes, and implications for experiment.
In the Appendix we provide some supplementary details.

\section{The Relativistic Schr\"{o}dinger Equation}\label{sec:relschro}

\subsection{Relativistic Hamiltonian}\label{sec:relham}
We first compute the Hamiltonian which governs the evolution of the environment probe particles in a curved spacetime. 
For convenience, we treat the probe particles as scalars, i.e., ignoring their spin. This is therefore not precise for realistic probe particles, such as baryons or photons, that can play the role of the environment (as we discuss later). However, one anticipates that the corrections from particle spin is small; we leave a proper analysis to possible future work. Then to incorporate coupling to gravity in a Lorentz invariant way, we use the field theoretic formalism for a quantum scalar field minimally coupled to gravity; this formalism makes it simple to describe interactions among particles in a local and unitary way.

Furthermore, we can take the scalar field to be complex for convenience. This is not important, since we will only be interested in single particle states, but it will make the formal manipulations simpler as we go on since it carries a conserved particle number.
For a minimally coupled complex scalar field $\scal$ in a general spacetime background with metric $g_{\mu\nu}$ (signature $+---$), the action for this field is 
\beq S = \int d^4 x \sqrt{-g}\left(g^{\mu\nu}\p_\mu \scal^*\p_\nu\scal - m^2 \scal^*\scal\right)
\eeq
where $g$ is the metric determinant, $g^{\mu\nu}$ is the inverse of the metric, $\scal^*$ is the complex conjugate of $\scal$, and $m$ is its mass (later we will use notation $m\to m_p$ to emphasize that it is a ``probe" particle). We ignore other possible interactions of $\scal$ here, since, as far as is known, particles may only couple to DM through gravity (we leave a discussion of other possible interactions to Section \ref{OtherInt}).

The momentum conjugate is given by
\beq 
\Pi \equiv \frac{\p\cL}{\p\dot{\scal}}=\sqrt{-g}\left(g^{00}\dot{\scal}^*+g^{0i}\p_i\scal^*\right) 
\eeq
The corresponding Hamiltonian density operator $\hat{\cH}$ is 
\beq 
\hat{\cH}=\frac{\Pi^*\Pi}{\rg g^{00}}-\frac{g^{0i}}{g^{00}}(\Pi\p_i\scal+\Pi^*\p_i\scal^*)+\frac{\rg g^{0i}g^{0j}}{g^{00}}\p_j\scal^*\p_i\scal-\rg g^{ij}\p_i\scal^*\p_j\scal+\rg m^2 \scal^*\scal
\eeq

Following the usual canonical quantization, where the fields and conjugate fields are promoted to operators and canonical commutation relations are imposed, we can then write out the operators in terms of a set of creation and annihilation operators, $a_\pv$ and $a_\pv^\dagger$, for the scalar particle described by the fields, and a set of similar operators, $b_\pv$ and $b_\pv^\dagger$, for its antiparticle -- these are summarized in Appendix \ref{FieldExpansion}. 
%\beq 
%\scal(\x)=\int \frac{ d^3 p}{(2\pi)^3}\frac{1}{\sqrt{2 E_p}}\left(\hat{a}_\pv e^{i\pv\cdot\x}+\hat{b}_\pv^\dagger e^{-i\pv\cdot\x}\right)
%\eeq
%\beq 
%\Pi(\x)=-i\int \frac{d^3p}{(2\pi)^3}\sqrt{\frac{E_p}{2}}\left(\hat{b}_\pv e^{i\pv\cdot\x}-\hat{a}_\pv^\dagger e^{-i\pv\cdot\x}\right)
%\eeq
%The factors of $E_p$, defined as $E_p\equiv\sqrt{p^2+m^2}$, assure the proper relativistic normalization, so that $\psi$ transforms as a Lorentz scalar, with standard creation and annihilation operators $[\hat{a}_\pv,\hat{a}^\dagger_{\pv'}]=\delta^3(\pv-\pv')$.
Single particle momentum eigenstates can be obtained from the vacuum state $\ket{0}$ as usual as
\beq 
\ket{\pv}=\sqrt{2E_p}\,\hat{a}_\pv^\dagger\ket{0}
\eeq

Then acting on the one-particle momentum eigenstate with the Hamiltonian operator 
\beq
\hat{H}\equiv \int d^3 x\, \hat{\cH}
\eeq 
gives the form of the Hamiltonian in momentum-space for this state. 

In this work, we focus on the situations in which one probe particle comes in, gravitationally scatters off the DM, and then this one probe particle goes out. There are, however, various other processes that can take place; this includes one probe particle comes in and three particles go out (if there is a conserved particle number, such as baryon number for protons as probe particles, then it would be two protons and one anti-proton going out). This is certainly allowed by general considerations (unless it is energetically forbidden for non-relativistic probes). But the physical point is the following: since we are studying the gravitational interaction, this would be suppressed by additional powers of Newton's gravitational constant $G_N$. In the language of Feynman diagrams, we are focussed on the simple tree process of $p+DM \to p + DM$
via a single graviton exchange. The scattering amplitude is proportional to $G_N$. However, we can also consider another tree process (or loop processes) in which $p+ DM \to p+p+\bar{p} + DM$ via 2 graviton exchange. This scattering amplitude is proportional to $G_N^2$. To fix the units there will be an additional factor of $E^2$, where $E$ is the probe particle's energy, in the amplitude. Assuming $G_NE^2\ll 1$ (i.e., for probes of energy much lower than the Planck energy) this is completely suppressed compared to the leading process. Hence states of fixed particle number are the dominant process to consider, which is what we focus on.

Furthermore, since the scatterers are assumed to be dilute, they will scatter one at a time. So we only need to consider the effect of a single probe at a time, compute its scattering amplitude, determine the overlap of wave-functions (which affects the density matrix), and then later combine the effects of waiting for $N$ scatterers which act essentially multiplicatively to the total wave-function. 

The Hamiltonian acting on a single particle state is
\ba
\hat{H}\ket{\q}=\int d^3 x\, \hat{\cH} \ket{\q}\amp= \amp\int d^3 x \frac{d^3 p}{2(2\pi)^3}\Bigg[\frac{\Eq }{\rg g^{00}}+\frac{g^{0i}}{g^{00}}\frac{p_i \Eq }{E_p}+\frac{g^{0i}}{g^{00}}q_i\nonumber\\
\amp\amp+\rg\left(\frac{g^{0i}g^{0j}}{g^{00}}-g^{ij}\right)\frac{p_iq_j}{E_p}+\rg\frac{m^2}{E_p}\Bigg]e^{-i(\pv-\q)\cdot\x}\ket{\pv}
\ea
Then, acting from the left with $\bra{\x'}$ we can write the Hamiltonian as a function of $\x'$ and derivatives in $\x'$ acting on a plane wave (the spatial wave function of the momentum eigenstate) as follows. Note that we only allow derivatives to act on the wave function and thus assume the metric components $g_{\mu\nu}$ to be sufficiently slowly varying in space. With $\langle{\x'}\ket{\pv}=e^{i\pv\cdot\x'}$ (note that we are using the compact notation $|{\bf x}'\rangle\equiv\scal^\dagger({\bf x}')|0\rangle$), we can then carry out the above integrals by performing some integration by parts. This leads to 
\beq
\bra{\x'}\hat{H}\ket{\q} = H(x',-\nabla'^2)e^{i\q\cdot\x'}
\eeq
where we have introduced a Hamiltonian $H$ defined in the position representation as a differential operator. It is found to be
\ba
H(x,-\nabla^2)\amp=\amp\frac{1}{2}\Bigg[\frac{\sqrt{-\nabla^2+m^2}}{\rg g^{00}}+2\frac{g^{0i}}{g^{00}}(-i\p_i)\nonumber\\
\amp+\amp \rg\left(\frac{g^{0i}g^{0j}}{g^{00}}-g^{ij}\right)\frac{-\p_i\p_j}{\sqrt{-\nabla^2+m^2}}+\rg\frac{m^2}{\sqrt{-\nabla^2+m^2}}\Bigg]
\ea
Note that even though we have built the relativistic theory using creation and annihilation operators in the usual way, we are allowed to act on particle states, and then form a kind of position space wave-function. Ordinarily, such position space wave-functions would couple multiple particle wave-functions to one another. However, as explained above, such multi-particle states are suppressed, which is to say, their amplitudes are small. This allows for a single particle wave-function to provide an accurate description of the state of the probes.

\subsection{Weak Diagonal Metric}\label{WDM}

We specialize now to a weakly curved spacetime. Furthermore, we assume that there are no significant sources of gravitational waves. Hence we can use a gauge in which the metric is diagonal. The metric can be decomposed into a flat background $\eta_{\mu\nu}=\textrm{diag}(+1,-1,-1,-1)$ and a small perturbation, $h_{\mu\nu}$, with $|h_{\mu\nu}|\ll1$, as follows
\beq 
g_{\mu\nu}=\eta_{\mu\nu}+h_{\mu\nu}
\eeq

Then, to linear order in the perturbations $h_{\mu\nu}$, the Hamiltonian becomes
\beq
 H(\x,t,-\nabla^2)  \approx \sqrt{-\nabla^2+m^2}+\frac{h^{00}}{2}\sqrt{-\nabla^2+m^2}-\frac{h^{ij}\p_i\p_j}{2\sqrt{-\nabla^2+m^2}}
\eeq
(writing $g^{\mu\nu}=\eta^{\mu\nu}-h^{\mu\nu}$). Furthermore, we will be focussed on metrics which are either static or, if they are time-dependent, are also spherically symmetric. In either case, we can fully specify the metric in terms of just two functions $\Phi$ and $\Psi$, which are related to components in the metric by
\beq 
h^{00}=2\Phi,\,\,\,\,\,\,\,\,\,\,
 h^{ij}=2\Psi \delta^{ij}
 \eeq
 Depending on the source for the metric, these two potentials $\Phi$ and $\Psi$ may or may not be the same. 
The corresponding Hamiltonian simplifies further to
\beq 
H(\x,t,-\nabla^2)=\sqrt{-\nabla^2+m^2}+\Phi\sqrt{-\nabla^2+m^2}-\frac{\Psi\nabla^2}{\sqrt{-\nabla^2+m^2}}
\eeq

Note that if we now take the nonrelativistic limit we obtain a familiar form
\beq 
H_{nr}(\x,t,-\nabla^2)=m\,c^2-{\nabla^2\over2m}+m\,\Phi
\eeq
where in the first term we reinstated a factor of $c^2$, to make it clear that this is merely the mass-energy term and is only a constant. However, we will only make use of the above relativistic Hamiltonian in this paper.

\subsection{One Particle Schr\"{o}dinger Equation}\label{sec:schro}

Given this relativistic Hamiltonian for a particle in a diagonal, weakly curved background spacetime, we can construct the Schr\"{o}dinger equation for the evolution of a quantum state that describes this particle. The Schr\"{o}dinger equation is of the usual form
\beq 
i \p_t\ket{\psi}=(\hat{H}_0+\hat{V})\ket{\psi}
\eeq
where in the position representation $H_0(-\nabla^2)=\sqrt{-\nabla^2+m^2}$ is the free theory relativistic Hamiltonian and the potential $V$ is defined as the remaining interaction parts of the Hamiltonian
\beq 
V(\x,t,-\nabla^2)\equiv\Phi(\x,t) \sqrt{-\nabla^2+m^2}-\frac{\Psi(\x,t) \nabla^2}{\sqrt{-\nabla^2+m^2}}\label{Vgen}
\eeq
The time dependence in eq.~(\ref{Vgen}) is made explicit because, in general, the potentials $\Phi$ and $\Psi$ can be time dependent.
The Schr\"{o}dinger equation for the wave function $\psi(\x,t)$ can be then be written as
\beq 
(i\p_t -\sqrt{-\nabla^2+m^2})\psi(\x,t)=V(\x,t,-\nabla^2)\psi(\x,t)
\eeq
Hence we establish a well defined single particle Schr\"{o}dinger equation, which can accurately describe processes in which particle number changing effects are negligible.

\section{Scattering and Decoherence in General}\label{ScatDecGeneral}

\subsection{Perturbative Expansion}

For weak gravitational interactions, it is useful to expand the full solution into a sum of an unscattered part $\psi_u$ and a scattered part $\psi_s$ as
\beq 
\psi(\x,t)=\psi_u(\x,t)+\psi_s(\x,t)
\eeq
The free particle wave function for a probe particle of mass $m_p$ (we replace $m\to m_p$ now) solves the free Hamiltonian Schr\"{o}dinger equation
\beq (i \p_t - \sqrt{-\nabla^2+\mpr^2})\psi_u(\x,t)=0 
\eeq
In fact we have $\psi=\psi_u$ at early times, before any interaction with the DMSCS. Then when encountering the DMSCS, the scattered part $\psi_s$ becomes nonzero. We can solve for the scattered part of the solution perturbatively by expanding in powers of the potential. Working to first order, the scattered part is given by the solution to the following equation
\beq 
(i\p_t-\sqrt{-\nabla^2+m_p^2})\psi_s(\x,t)=V(\x,t,-\nabla^2)\psi_u(\x,t)
\eeq
where $\psi_s$ will from now on denote the first order contribution to scattering and not the full scattering solution; working to first order will be sufficient to determine the decoherence rate to leading order (for further details on this point; see Ref.~\cite{Allali:2020ttz}). Having demanded that $\psi_s\to0$ at early times, the solution for $\psi_s$ is provided by the particular solution to this equation 
\beq\psi_s(\x,t)=\int d^4x' G_4(t-t',\x-\x') V(\x',t',-\nabla'^2)\psi_u(\x',t')\eeq
where $G_4$ is the four-dimensional (time-dependent) retarded Green's function of the time-dependent Schr\"{o}dinger operator
\beq 
(i\p_t-\sqrt{-\nabla^2+m_p^2})G_4(t-t',\x-\x')=\delta(t-t')\,\delta^3({\bf x}-{\bf x}')
\eeq

Since $\psi_u$ is just the solution to the free equation, the $k^{\text{th}}$ mode is both a momentum and energy eigenstate, and thus its position representation wave function $\psi_u^{(k)}$ has the simple time dependence given by
\beq \psi^{(k)}_u(\x,t)= e^{-i \Ek  t} \psi^{(k)}_u(\x) \eeq
where $\psi^{(k)}_u(\x) = e^{i\k\cdot\x}$ (with some normalization) is the spatial part of the wave function of the $k^{\text{th}}$ mode, $\Ek =\sqrt{k^2+m_p^2}$ is its energy, and $\psi^{(k)}_u(\x,t)$ is its full time-dependent wave function.

\subsection{Time-Independent Green's Function}

In the upcoming sections we will study two cases: (i) static DMSCS sources and (ii) time-dependent DMSCS sources. In the first case (i) we will be able to solve the problem perturbatively using a spatial (three-dimensional) Green's function. In the second case (ii) the time dependence will be more complicated, requiring use of the full space-time dependent Green's function. However, we will focus on coherently oscillating DMSCS sources, which carry a simple harmonic time dependence. As we will show, this will allow us to still solve the problem in terms of an appropriately shifted spatial Green's function.

The three-dimensional (spatial) Green's function is found by taking the free theory ($V=0$) time-independent Schr\"{o}dinger equation, Fourier transforming in time, and inserting a three-dimensional delta function as a source
\beq 
(E-\sqrt{-\nabla^2 +m_p^2}) G_3(E, \x-\x')=\delta^3(\x-\x')
\eeq
To obtain this Green's function, we first take the spatial Fourier transform
\beq \tilde{G}_3(E,\pv)=\int d^3 x\, G_3(E, \x-\x') \,e^{i\pv\cdot(\x-\x')}
\eeq
The corresponding particular solution is
\beq 
\tilde{G}_3(E,\pv) = \frac{1}{E-\sqrt{p^2+m_p^2}+i\epsilon}
\eeq
where the $i\epsilon$ factor is chosen to obtain the retarded Green's function. Defining $\rv=\x-\x'$, we can readily obtain the Green's function through the inverse Fourier transform. We can easily carry out the angular integral to obtain
\beq
 G_3(E, \x-\x') =\frac{1}{2\pi^2}\int^\infty_0 dp\, p^2 \,\frac{\sin pr}{pr}\frac{1}{E-\sqrt{p^2+m_p^2}+i\epsilon}
\eeq
Changing to $\en\equiv\sqrt{p^2 +m_p^2}$ such that $p = \sqrt{\en^2-m_p^2}$ and $\en \, d\en = p\, dp$ and rewriting $\sin pr$ in terms of complex exponentials, this can be rewritten as
\beq 
G_3(E,\x-\x') =-\frac{1}{2\pi^2 r}\int^\infty_{m_p} \frac{e^{i p r}-e^{-ipr}}{2 i}\frac{\en \,d\en}{\en- (E+i\epsilon)}
\eeq
where the remaining $p$ dependence is implicitly dependent on $\en$.
Integration may now be done in the complex plane. For convergence, a contour around the upper half of the plane is chosen for the term proportional to $e^{ipr}$, and thus the pole at $\en\rightarrow E+i\epsilon$ contributes to the integral. Similarly, the contour in the lower half of the plane is chosen for the term proportional to $e^{-ipr}$ and thus, lacking a pole, there is no contribution to the integral. At large radius, which is all we will need later on, we can ignore the branch cut starting at $\en=\pm m_p$. 
We are thus left with
\beq G_3(E, \x-\x')=-\frac{1}{2\pi^2 r} 2\pi i \,\textrm{Res}_{\en\rightarrow E+i\epsilon}\left(\frac{e^{ipr}}{2i(\en-(E+i\epsilon))}\right)=-\frac{1}{2\pi r}e^{ipr} E \eeq
where in the last step we have taken $\epsilon\rightarrow 0$. Note that the remaining $p$ in the formula is implicitly defined from the energy $E$ as $p=\sqrt{E^2-m_p^2}$ (as a result of the $d^3p$ integration). As expected, in the nonrelativistic limit $E\to m_p$, we recover the usual time-independent Schr\"{o}dinger Green's function. 

Also, we note that if one begins from the Klein-Gordon equation (see Appendix \ref{KGapp}) and uses the corresponding Feynman propagator, one of course has a pair of poles. One pole gives the above contribution, while the other pole is highly off-shell for positive energy particles and is ignorable (as it is related to anti-particles), so together we recover the Green's function above.

\subsection{Wave Packets}\label{sec:wavepckt}

We will assume the incoming wave function of the probe particle is a wave packet. We can take the incoming, unscattered wave to be a sum of plane wave modes with a distribution function $\psik(\q)$ which weights the momenta $\q$ of the modes around a central value $\k$ which can be the thought of as the mean momentum. Hence at early times, the spatial wave function is given by
\beq 
\psi_u(\x)=\int \!\frac{d^3q}{(2\pi)^3}\frac{1}{\sqrt{2\Eq }}\psik(\q) e^{i\q\cdot\x}e^{-i\q\cdot\bv}
\eeq
where the vector $\bv$ is an impact parameter which can shift the spatial center of the wave packet away from the origin of the coordinate system. The pre-factor $1/\sqrt{2 E_q}$ is convenient, because the normalization condition $\langle\psi_u|\psi_u\rangle=1$ takes on the simple form
\beq 
\int \!\frac{d^3q}{(2\pi)^3}|\psik(\q)|^2=1 
\eeq
(see ahead to eq.~(\ref{overlap}) for the general expression for the overlap between a pair of 1-particle states).

\subsection{Decoherence }

\begin{figure}[t]
\centering
\includegraphics[width=\columnwidth]{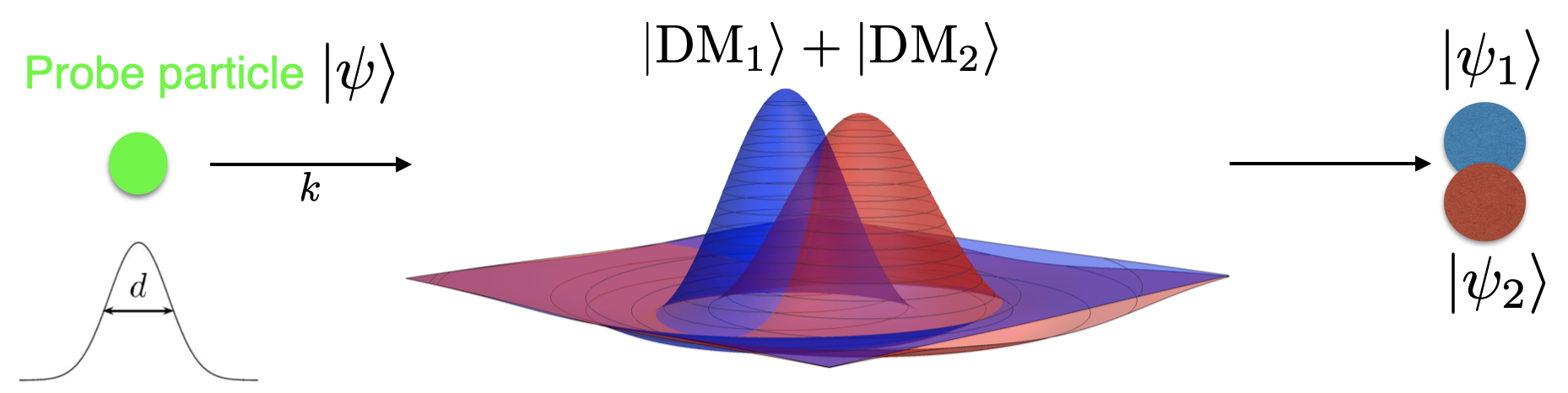}
\caption{The basic setup of the problem; this figure is taken from our previous paper Ref.~\cite{Allali:2020ttz}. 
Suppose there is some piece of matter that primarily only interacts gravitationally, dark matter, and it forms a dark-matter-Schr\"{o}dinger-cat-state (DMSCS). This is defined as a quantum superposition of two different configurations of some piece of the DM,  $\ket{\DM_1}+\ket{\DM_2}$, as illustrated by the blue and red distributions. The distributions can differ in the center of mass position and/or profile of the mass distribution or, for light bosonic DM, the phase of the oscillating DM field.
 Now also suppose there is a probe particle (wave packet with wavenumber $k$ and width $d$) in state $\ket{\psi}$ which passes through; its gravitational interaction with the DMSCS ensures that it evolves into a superposition of states $\ket{\psi_1}$ and $\ket{\psi_2}$, and so it is then entangled with the DM.}
\label{fig:setup}
\end{figure}

Our interest will be a DMSCS that begins in a superposition of otherwise classical states $|\mbox{DM}_1\rangle$, $|\mbox{DM}_2\rangle$. Using the Einstein equations of general relativity (for details see later in the paper), this sets up an effective potential $V$ that is itself in a superposition of potentials $V_1$ and $V_2$; see Fig.~\ref{fig:setup}. The DM can be in a superposition of different density profiles and, if time dependent, in different phases, etc. For further generality, we also allow the centers of the potentials (i.e., the center of the mass distribution) to differ from the origin of the coordinate system, characterized by the distance $\Lv_i$ (here the index $i$ is either $1$ or $2$, denoting correspondence with either $V_1$ or $V_2$). 
The evolution of these environmental probe particles is characterized by a wave function, $\psi$, evolving according to the Schr\"{o}dinger equation. After interacting with the potential ($V_1$ or $V_2$), the probe particle also evolves into a superposition of sub-states $|\psi_1\rangle,\,|\psi_2\rangle$, which are inevitably entangled with the DMSCS. The total final state is given by
\beq
|\Psi_f\rangle=|\psi_1\rangle\,|\mbox{DM}_1\rangle+|\psi_2\rangle\,|\mbox{DM}_2\rangle
\eeq
By tracing out these probe particles, which act as the ``environment" that we do not track closely, the reduced density matrix of just the dark matter describes a mixed state. Our goal then is to compute the rate at which interactions with such particles lead to decoherence of the DMSCS. 

As is known (e.g., see Ref.~\cite{Allali:2020ttz}), the decoherence rate depends on the inner product of the sub-states  $|\psi_1\rangle,\,|\psi_2\rangle$. We parameterize the overlap of these two sub-states in terms of $\Delta$ as
\beq 
|\braket{\psi_1}{\psi_2}|\equiv 1-\Delta 
\eeq
To be consistent with the relativistic normalization of the quantum states defined in Section~\ref{sec:relham}, the inner product in the position representation of a generic pair of 1-particle states is
\beq
\braket{\psi}{\phi}\equiv\int d^3 x \left(-i \phi(\x,t)\p_t \psi^*(\x,t)+i\psi^*(\x,t)\p_t\phi(\x,t)\right) \label{overlap}
\eeq

The deviation from unity $\Delta$, to lowest order in scattering, can be readily shown to be given by
\beq 
\Delta=\frac{1}{2}\left( \braket{\psi_{s,1}}{\psi_{s,1}}+\braket{\psi_{s,2}}{\psi_{s,2}}-2\Re\left[\braket{\psi_{s,1}}{\psi_{s,2}}\right]\right)\label{Delta}\eeq
The decoherence rate is then given by (see our Ref.~\cite{Allali:2020ttz} for details)
\beq
\Gamd=n\,v\int \! d^2b\,\Delta_b
\label{DecRate}\eeq
where $n$ is the number density of probe particles, $v$ is their typical speed, and $b$ is the impact parameter of each probe's approach, which is integrated over to account for its variation.

\section{Static Source }\label{sec:static}

We now wish to analyze the evolution of an environment particle in the spacetime generated by the DMSCS. First we examine the case where the background spacetime produced by the DMSCS is static $V=V(\x,-\nabla^2)$. The evolution of the environment particles is therefore described as the scattering of a probe particle by a fixed potential. We assume the DMSCS is taken to be sufficiently massive such that the backreaction is negligible.

For a truly nonrelativistic DMSCS source, the pressure of the source should be negligible, and thus the potentials $\Phi$ and $\Psi$ are equal. In fact, for a DMSCS that is an oscillating scalar field, like the axion, taking the source to be static should correspond to this nonrelativistic limit. However, to generalize the following results to a source which is not an oscillating scalar, we will keep the contribution of the source's pressure, and thus continue to differentiate between $\Phi$ and $\Psi$. This may be useful to estimate the decoherence of an object which is relativistic but still sources a static metric.

\subsection{Scattered Wave Function for Static Case}

Since the source is static, it is useful to relate the four-dimensional and three-dimensional free Green's functions in a simple way due to the separability of the free Schr\"{o}dinger equation
\beq G_3(E,\x-\x')=\int dt' G_4(t-t',\x-\x')e^{-i E (t'-t)} \eeq
Thus the scattered response of the $k^{\text{th}}$ mode may be written as
\begin{align}
\psi_s^{(k)}(\x,t)
&=e^{- i \Ek  t} \int d^3 x' G_3(\Ek ,\x-\x') V(\x',-\nabla'^2) \psi^{(k)}_u(\x)\nonumber\\
&=e^{- i \Ek  t}\int d^3 x' \left(\frac{- \Ek  e^{i k |\x-\x'|}}{2\pi |\x-\x'|}\right)\left(\Phi(\x') \sqrt{-\nabla'^2+m_p^2}-\frac{\Psi(\x') \nabla'^2}{\sqrt{-\nabla'^2+m_p^2}}\right)\psi^{(k)}_u(\x)\label{bigint}
\end{align}

We will now focus on the far distance regime $|\x|\gg|\x'|$. This is of most importance at late times when the wave function is concentrated far from the center of the potential. In the exponential phase factor, we can expand $|\x-\x'|\approx |\x|-\hat{x}\cdot\x'$. Also, given that the wave function of the $k^{\text{th}}$ mode is a plane wave of momentum $\k$, with $\psi_u^{(k)}(\x)= e^{i\k\cdot\x}$, the derivatives in the potential can act on this wave function, giving the following expression for the scattered response of the $k^{\text{th}}$ mode
\begin{align}
\psi_s^{(k)}(\x,t)
&=e^{- i \Ek  t}f(\k',\k)\frac{e^{ik|\x|}}{|\x|}
\end{align}
where $\k'\equiv k \hat{x}$ and we have defined the first order scattering amplitude as
\beq 
f(\k',\k)\equiv -\frac{1}{2\pi}\int d^3 x' e^{i(\k-\k')\cdot\x'}\left[\Phi(\x')(k^2+m_p^2)+\Psi(\x')k^2\right]\label{staticamp}
\eeq
Then, the wave function for the scattered wave packet is given by
\beq 
\psi_s(\x,t) = \int \frac{d^3q}{(2\pi)^3}e^{-i\Eq  t}f(\q',\q)\frac{e^{iq|\x|}}{|\x|}\frac{ \psik(\q)}{\sqrt{2\Eq }}e^{-i\q\cdot\bv}
\eeq
This generalizes the usual nonrelativistic treatments, e.g., see Refs.~\cite{Sakurai,MurayamaWebpage} (we will later provide an even much greater generalization to the time-dependent relativistic case).

\subsection{Decoherence Rate}\label{sec:static decrate}

We now apply the above result for the scattered wave function to obtain a decoherence rate. The DMSCS is in a superposition, meaning that when the probe wave-packet passes by it will itself launch into a superposition. The scattered parts of the superposition will be denoted $\psi_{s,i}$ or $\psi_{s,j}$, where $i,j=1,2$ for each part of the superposition.

In order to obtain the decoherence rate, the following integral over impact parameter is of interest
\beq
S_{ij}= \int d^2 b \braket{\psi_{s,i}}{\psi_{s,j}}\label{SijDef}
\eeq
Using the above result for $\psi_s$, we see that this is given by
\ba 
S_{ij}\amp=\amp\int d^2 b \int d^3 x \frac{d^3 q d^3 \qt}{(2\pi)^6}\left[ e^{-i E_{\qt}t}f_j(\qtv',\qtv)\frac{e^{i\qt r_j}}{r}\frac{\psik(\qtv)}{\sqrt{2E_{\qt}}}e^{-i\qtv\cdot(\bv-\Lv_j)}\right]\nonumber\\
\amp\amp\times( E_q+E_{\qt})\left[ e^{+i E_q t}f_i^*(\q',\q)\frac{e^{-iq r_i}}{r}\frac{\psik^*(\q)}{\sqrt{2\Eq }}e^{+i\q\cdot(\bv-\Lv_i)}\right]
\ea
Here we have defined $r_i\equiv |\x-\Lv_i|\approx r - \hat{x}\cdot\Lv_i$. We can perform parts of the integral as follows: Firstly the integral over impact parameter gives a two-dimensional delta function
\beq
\int d^2 b\, e^{-i\bv\cdot(\q-\qtv)}=(2\pi)^2\delta^2(\q_\perp - \qtv_\perp)\\\label{bint}
\eeq
while the integral over space can be written in spherical co-ordinates, and the radial integral gives another delta function
\beq
\int d^3 x\frac{e^{-i(q-\qt)r}}{r^2}=\int d^2\Omega\int ^\infty_0 dr \, e^{-i(q-\qt)r}=\int d^2\Omega (2\pi) \delta(q-\qt)\label{rint}
\eeq
The combination of these delta functions results in two possibilities $\q=\qtv$ and $\q=-\qtv$. However, we can use the fact that the distributions $\psik(\q)$ are sufficiently narrowly peaked such that the case when $\q$ and $\qtv$ are anti-aligned is exponentially suppressed. Thus we approximate the delta functions $(2\pi)^3\delta^2(\q_\perp-\qtv_\perp)\delta(q-\qt)\approx(2\pi)^3\delta^3(\q-\qtv)$ and perform the $d^3\qt$ integrals, resulting in
\beq
S_{ij}=
\int d^2\Omega\int \frac{d^3q}{(2\pi)^3}f_i^{*}(\q',\q)f_j(\q',\q)e^{-i(\q-\q')\cdot(\Lv_i-\Lv_j)}|\psik(\q)|^2\label{Sijstatgen}
\eeq

At this point, it can be clearly seen that the results of these integrals can follow the same calculations as in the nonrelativistic Newtonian case in Ref.~\cite{Allali:2020ttz}, with the exception of the fact that the scattering amplitudes $f$ are defined differently for the relativistic case. It is useful to average over the direction of $\k$ (as in done in the previous work), which impacts the exponentials in eq.~(\ref{Sijstatgen}) and replaces $|\psik(\q)|^2$ with a function $P_k(q)$ which depends only on the magnitudes $k$ and $q$. It is then convenient to define the following generalized cross section which reflects the averaging over the direction of $\k$
\beq
\sigmat_{ij}(q)\equiv\int \!d^2\Omega\,f_i^{*}(\q',\q)f_j(\q',\q)\,j_0(2q L_{ij}\sin(\theta/2))\label{sigmatilde}
\eeq
\beq S_{ij}=\int \frac{d^3 q}{(2\pi)^3}\sigmat_{ij}(q)P_k(q) \eeq
where $j_0(z)\equiv \sin(z)/z$ is the sinc-function and $L_{ij}\equiv|{\bf L}_i-\Lv_j|$. Note that $\sigma_1\equiv\sigmat_{11}$ and $\sigma_2\equiv\sigmat_{22}$ correspond to the usual definition of a scattering cross section (with the appropriate general relativistic amplitudes of eq.~(\ref{staticamp})) since $j_0(0)=1$. Finally, the decoherence rate is given by
\beq 
\Gamd= \frac{1}{2} n v (S_{11}+S_{22}-2\Re[S_{12}])
\eeq
This may be approximated by taking $S_{ij}$ to be given by the $\tilde{\sigma}_{ij}$ evaluated at $q=k$; this approximation is valid if the distributions $\psik$ are sufficiently narrow such that the distribution $P_k(q)$ is sharply peaked at $q=k$. Then we can express the decoherence rate as
\beq 
\Gamd= \frac{1}{2} n\, v (\sigma_{1}+\sigma_{2}-2\Re[\sigmat_{12}])|_{q=k}\label{GamdecStatic}
\eeq
Note that in the special limit in which $\tilde{\sigma}_{12}=0$ and $\sigma\equiv\sigma_{1}=\sigma_{2}$, this simplifies to the familiar form $\Gamd=n\,v\,\sigma$ relating to other work in the literature on decoherence; but here it is generalized to nontrivial overlap (provided by the $\tilde{\sigma}_{12}$ term) and to a general relativistic setting. 

\subsection{Parameterization of the Decoherence Rate}\label{ParStatic}

To compute the decoherence rate explicitly, one must now provide a specific form for the potentials $\Phi$ and $\Psi$. For a static, diagonal, spherically symmetric metric, one can readily show that the Einstein equations imply $\Psi=\Phi_N$, $\Phi=\Phi_N+\delta$, where $\Phi_N$ is the Newtonian gravitational potential and $\delta$ is a correction
\beq
\nabla^2\Phi_N = 4\pi G_N\, \rho \label{poissonstat},\,\,\,\,\,\,\,\,\,\,
{1\over r}\,\delta' = 4\pi G_N\,\press_r
\eeq
where $\rho$ is the enegy density of the source and $\press_r$ is the radial pressure. Most sources lead to a potential that has ``infinite range'', leading to a divergence of the scattering cross section in the forward scattering direction (e.g. Rutherford scattering formula for a Coulomb potential). This divergence may be avoided for mass distributions with a vanishing monopole.

One important mass distribution to consider for DM in the galaxy is a fluctuation in the background density of DM which is an overdensity surrounded by an underdensity such that the total perturbation in the integrated mass (monopole) from the background vanishes. We can parametrize this profile with (a) a characteristic mass scale, $M$, which can be thought of as the mass scale associated with the overdensity (since the overdensity and underdensity together have no mass); and (b) a characteristic length scale $1/\mu$, which can be thought of roughly as the width of the region including the overdensity and underdensity, and which we will take to be of the order of the de Broglie wavelength of the DM. 
Then, the mass density that deviates from the background of DM is given by 
\beq 
\delta\rho(\x)\equiv M\mu^3\zeta(\mu r),\,\,\,\,\,\,\,\delta \press_r(\x)\equiv M\mu^3\gamma(\mu r)
\eeq
where $\zeta(\mu r),\,\gamma(\mu r)$ are dimensionless functions which describe the shape of the energy density and pressure of the sources, respectively, and depend on the dimensionless variable $\mu r$. Note that we may take the distribution to be spherically symmetric for simplicity, and thus the sources only depends on the position variable $|\x|=r$. 

The scattering amplitude in eq.~(\ref{staticamp}) is related to the Fourier transform of the potential with respect to the transfer momentum $\pv_{tr}=\k-\k'$. We can define the dimensionless variables $\hat{\x}\equiv \mu \x$ and $\hat{\pv}\equiv\frac{\pv_{tr}}{\mu}$ (with $|\hat{\pv}|=\hat{p}=\frac{2k}{\mu}\sin(\theta/2)$ and $\theta$ the angle between $\k$ and $\k'$) to have a dimensionless Fourier transform as follows
\ba
\hat{F}({\bf{\dv}})\amp\equiv\amp\int d^3\hat{x} \,F(\y) \,e^{i\bf{\dv}\cdot\y}\label{fourierdef}
\ea
The scattering amplitude is then simply given by
\beq 
f(\k',\k)=-\frac{1}{2\pi\mu^3}\left[\hat{\Phi}_N(\hat{p})(m_p^2 +2k^2)+\hat{\delta}(\hat{p})(m_p^2 +k^2)\right]
\eeq
Then, taking the Fourier transform of eq.~(\ref{poissonstat}), we can write down a form of the scattering amplitude with the physical parameter dependence more clearly displayed
\beq 
-\dv^2\,\hat{\Phi}_N(\dv)=4\pi G_N\, M\,\mu\, \hat{\zeta}(\dv)
\label{PoissonFT}\eeq 
Similarly, to find a Fourier transform for $\delta$, we may first rewrite eq.~(\ref{poissonstat}) for $\delta$ by differentiating once with respect to $r$ and utilizing the fact that we have assumed that $\delta$ is spherically symmetric
\beq 
\nabla^2\delta=4\pi G_N M \mu^3 \Big(3\gamma(\mu r)+\mu r \gamma'(\mu r) \Big)
\eeq
Then, defining the function $\varepsilon$ as 
\beq \varepsilon(\mu r)\equiv 3\gamma(\mu r) +\mu r \gamma'(\mu r)\eeq
we can find the Fourier transform of $\delta$ from the following
\beq -\dv^2\,\hat{\delta}(\dv)=4\pi G_N\, M\,\mu\, \hat{\varepsilon}(\dv)
\eeq
Then finally, we can find the generalized cross sections of eq.~(\ref{sigmatilde}) to have the form
\beq 
\tilde{\sigma}_{ij}(k)
=\frac{8\pi G_N^2 M_i M_j}{\mu_i\mu_j}\frac{(m_p^2+2k^2)^2}{k^2} \chi_{ij}
\eeq
where we have defined a dimensionless quantity
\beq
\chi_{ij}\equiv \int \frac{d \dv}{\dv^3}\, \left[\hat{\zeta}\!\left(\sqrt{\frac{\mu_i}{\mu_j}}\dv\right)
+\left(\frac{m_p^2+k^2}{m_p^2+2k^2}\right)\hat{\varepsilon}\!\left(\sqrt{\frac{\mu_i}{\mu_j}}\dv\right)\right]
\times\left[i\leftrightarrow j\right]\times
j_0(L_{ij}\sqrt{\mu_i\mu_j}\,\dv)\label{sigmatilde2}
\eeq
and $\left[i\leftrightarrow j\right]$ means to repeat the previous quantity in square brackets, interchanging $i$ and $j$.

Having scaled out the various parameters, this remaining dimensionless quantity $\chi_{ij}$ is typically $\cO(1)$ (e.g., when the ratio of $\mu_i$ and $\mu_j$ is $\cO(1)$, $L_{ij}$ is small compared to $1/\mu$, and the potential is exponentially damped outside the region of radius $1/\mu$, like a Gaussian). Using eqs.~(\ref{GamdecStatic},\,\ref{sigmatilde2}) the decoherence rate is then given by
\beq 
\Gamd \approx 4\pi G_N^2 n_p v_p\frac{(m_p^2+2k^2)^2}{k^2}\left(\frac{M_1^2}{\mu_1^2}\ch_{11}+\frac{M_2^2}{\mu_2^2}\ch_{22}-\frac{2M_1M_2}{\mu_1\mu_2}\ch_{12}\right)
\eeq
Note the two features of relativity here: in the $(m_p^2+2k^2)^2$ prefactor and in the definition of $\chi_{ij}$ in eq.~(\ref{sigmatilde2}), which includes the $\hat\varepsilon$ term from pressure; this provides a neat extension of Ref.~\cite{Allali:2020ttz}.

To examine this further, let us now examine this result in the nonrelativistic ($k\ll m_p$) and ultrarelativistic ($k\gg m_p$) limits. In the nonrelativistic limit, we recover the result of our previous nonrelativistic analysis in Ref.~\cite{Allali:2020ttz}, namely
\beq 
\Gamma_{\mbox{\tiny{dec,nr}}} \approx 4\pi G_N^2 n_p \frac{m_p^2}{v_p}\left(\frac{M_1^2}{\mu_1^2}\ch_{11}+\frac{M_2^2}{\mu_2^2}\ch_{22}-\frac{2M_1M_2}{\mu_1\mu_2}\ch_{12}\right)\,\,\,\,\,\,(\mbox{nonrelativistic probes})
\eeq
A good example of nonrelativistic probe particle are the baryons in the galaxy, or in atmosphere of earth, which are plentiful. 
For a detailed analysis of this result applied to axionic (or other light bosonic) DM in the galaxy today, including estimates in the atmosphere of the earth and dilute boson stars, see the plots and discussion in Ref.~\cite{Allali:2020ttz}.

In contrast, the ultrarelativistic limit leads to different predictions
\beq 
\Gamma_{\mbox{\tiny{dec,ur}}} \approx 16\pi G_N^2 n_p k^2 \left(\frac{M_1^2}{\mu_1^2}\ch_{11}+\frac{M_2^2}{\mu_2^2}\ch_{22}-\frac{2M_1M_2}{\mu_1\mu_2}\ch_{12}\right)\,\,\,\,\,\,(\mbox{ultrarelativistic probes})
\eeq
where we take $v_p\to 1$ in the ultrarelativistic limit. A good example of ultrarelativistic probes are cosmic microwave background (CMB) photons. Here, we can see that the decoherence rate now increases with increasing probe particle momentum. Thus, one may expect that sufficiently high-energy particles will cause decoherence of the DMSCS to come about more rapidly. However, when considering DM in the galaxy, the relative abundance of high-speed particles compared to low-speed particles is quite low. Therefore, one should expect that, for example, nonrelativistic hydrogen in the galaxy will dominate the calculation of the decoherence rate.

To compare $\Gamma_{\mbox{\tiny{dec,nr}}}$ and $\Gamma_{\mbox{\tiny{dec,ur}}}$ directly amounts to comparing the factors $n_p m_p^2/v_p$ of nonrelativistic probe articles to $n_p k^2$ of ultrarelativistic probe particles. In the former, the larger abundance of probe particles should more than compensate for the fact that in the latter $k\gg m_p$. One way, however, to consider a probe which is generically ultrarelativistic is to consider a photon probe. In this case, it is still clear that the nonrelativistic hydrogen with $m_p^2/v_p \sim 10^3 \, \mbox{GeV}^2$ will dominate over the contribution of photons which either have much smaller energies or, if they are energetic enough to compete, have much lower abundances.

Therefore, it is reasonable to conclude that the nonrelativistic analysis is sufficient to compute the decoherence rate for static DMSCS configurations. This provides extra motivation to consider DMSCS which vary in time; these sources necessarily require a general relativistic treatment and are the subject of the next section. 

\section{Time Dependent Source}\label{sec:timedep}

We now consider the case where the metric sourced by the DMSCS varies in time. Of particular interest is light bosonic DM, such as an axion, which can oscillate in time and thus source a time-varying spacetime.

\subsection{Einstein Equations for Spherically Symmetric Source}

First, we can compute the background metric sourced by the energy momentum tensor of a scalar field which oscillates coherently in time. We consider the Lagrangian for the scalar field $\phi$ of mass $\ma$ (for simplicity we ignore self-interactions, like $\sim\lambda\phi^4$ in this discussion)
\beq 
\mathcal{L}=\frac{1}{2}\partial_\mu \phi \partial^\mu \phi - {1\over2}\ma^2 \phi^2
\eeq

From this, we construct the energy momentum tensor, which we only need to zeroth order in the gravitational field
\beq T_{\mu\nu} = \partial_\mu\phi\partial_\nu \phi -\eta_{\mu\nu}\mathcal{L}
\eeq
Assuming spherical symmetry $\phi=\phi(r,t)$, using notation $\partial_r\phi\equiv\phi'$, $\partial_t\phi\equiv\dot{\phi}$, and using spherical co-ordinates, we have
\ba
\rho\equiv T_{tt} \amp=\amp 
\frac{1}{2}(\dot{\phi}^2 + (\phi')^2+\ma^2\phi^2\label{EnDensity}\\
\press_r\equiv  T_{rr} \amp=\amp \frac{1}{2}(\dot{\phi}^2+(\phi')^2-\ma^2\phi^2) \\
\press_\theta\equiv {T_{\theta \theta}\over r^2} \amp=\amp \frac{1}{2}(\dot{\phi}^2-(\phi')^2-\ma^2\phi^2)
\ea
(and $T_{\phi\phi} = T_{\theta\theta}\sin^2\theta$). For convenience we will express the metric in spatially conformally flat coordinates as
\beq
g_{\mu\nu}=(1+2\Psi(r,t)+2\delta(r,t))dt^2-(1-2\Psi(r,t))(dr^2+r^2 d\theta^2+r^2\sin^2\theta d\varphi^2)\label{metricSCF}
\eeq
with $\Psi = \Psi(r,t)$ and $\Phi=\Psi(r,t)+\delta(r,t)$. 
From this metric, we can use the linearized Einstein equations $G_{\mu\nu}=8\pi G_N\, T_{\mu\nu}$ to solve for the metric as
\ba
(G_{tt}=)\,\,\,\,\,\,\,\,\,\,\,2\nabla^2\Psi\amp=\amp4\pi G_N(\dot\phi^2+(\phi')^2+\ma^2\phi^2)\\
(G_{\theta\theta}/r^2-G_{rr}=)\,\,\,\,\,\, r\!\left(\delta'\over r\right)'\amp=\amp-8\pi G_N(\phi')^2\label{deltaEqn}
\ea
Using the Klein-Gordon equation of motion for the scalar field, and the equations for $\Psi,\,\delta$ above, it can be shown that the $G_{tr}=8\pi G\,T_{tr}= 8\pi G_N\,\dot{\phi}\phi'$ equation is automatically satisfied, so we do not need to specify it here.

\subsection{Coherently Oscillating Scalar Field}

From these equations, we can calculate the metric sourced by an oscillating scalar. We will take the scalar field to have a spatial profile $\scprof(r)$ that is spherically symmetric and time dependence which is a single simple harmonic of angular frequency $\omega$. For diffuse DM that is in motion in the galaxy, this is not precise, since it is really made out of a combination of traveling waves, with some fluctuations in frequencies. However, for such diffuse DM, the variation in frequency is on the order $\omega=m_a+{1\over2}m_av^2$, where $v^2$ is the typical dispersion in velocities. For nonrelativistic DM, this correction is tiny (we shall return to these details in Section \ref{sec:disc}), so it is fairly monochromatic. Furthermore, for a condensate of scalars, one usually has almost perfectly periodic oscillations, so this form is even more appropriate (we shall return to such condensates in Section \ref{sec:outlook}).

In any case, for our discussion it will suffice to treat the scalar field as having the form
\beq \phi(r,t) = \sqrt{2}\,\scprof(r) \cos(\oma t+\varphi)
\eeq
Hence $\phi' = \sqrt{2}\,\scprof'(r)\cos(\oma t+\varphi)$ and $\dot{\phi}=-\sqrt{2}\,\oma \scprof(r)\sin(\oma t+\varphi)$. This can be inserted into eq.~(\ref{EnDensity}) to obtain the corresponding energy density. By using double angle formulas, it can be readily shown that it takes on the following form
\beq
\rho(r,t)= 
\rhoA(r)+\rhoB(r)\cos(2(\omega t+\varphi)) \label{rhocos2}
\eeq
where we have indicated that the energy density $\rho$ has a part which is independent of time and a part which is proportional to $\cos(2(\oma t+\varphi))$. The coefficients are functions of position given by
\ba
&&\rhoA(r)=\frac{\scprof^2 (\ma^2 - \oma^2)+(\scprof')^2}{2}+\scprof^2\oma^2 \\
&&\rhoB(r)=\frac{\scprof^2 (\ma^2 - \oma^2)+(\scprof')^2}{2}
\ea 
We can then identify that the Newtonian potential $\Psi$ has the same time dependence structure
\beq
\Psi(r,t)=\PsiA(r)+\PsiB(r)\cos(2(\omega t+\varphi))
\eeq
with each piece solving the corresponding Poisson equation
\beq
\PsiAB=4\pi G_N{\rhoAB\over\nabla^2}
\eeq
for $n=0,\,2$.
In addition, we can solve for $\delta$ from eq.~(\ref{deltaEqn}), from which we can see that $\delta$ will also be proportional to $\cos^2(\oma t+\varphi)$, and thus will also have a part independent of time and a part that oscillates in time with $\cos(2(\oma t+\varphi))$ as
\beq
\delta(r,t)=\deltaA(r)+\deltaB(r)\cos(2(\omega t+\varphi))
\eeq
with $\delta^{(0,2)}$ obeying
\beq
\deltaA(r)=\deltaB(r) = -8\pi G_N\int_r^\infty dr'\,r'\int_{r'}^\infty dr''\,{1\over r''}(\scprof')^2
\eeq
Thus, both potentials in the metric, $\Psi$ and $\Phi=\Psi+\delta$, have some part which is independent of time and some other part which depends on $\cos(2(\oma t+\varphi))$. 

\subsection{Scattered Wave Function for Dynamic Case}\label{sec:timevarscat}
We then wish to analyze the response of a one-particle wave function for a probe particle of mass $m_p$ evolving according to the relativistic Schr\"{o}dinger equation in this spacetime background. We change from describing the potentials $\Phi$ and $\Psi$ in favor of an equivalent set of time-independent potentials $\A(r,-\nabla^2)$ and $\B(r,-\nabla^2)$ in order to explicitly separate out the time dependence of $V$ from eq.~(\ref{Vgen})
\beq
V(x,-\nabla^2)
=\A(r,-\nabla^2)+\B(r,-\nabla^2)\cos(2(\oma t+\varphi))\label{potlcos2m}
\eeq
where we have defined the spatial operators
\beq
\AB(r,-\nabla^2) = (\PsiAB(r)+\deltaAB(r))\sqrt{-\nabla^2+m^2}-{\PsiAB(r)\nabla^2\over\sqrt{-\nabla^2+m^2}}
\eeq
for $n=0,\,2$.
Then, we solve the Schr\"{o}dinger equation perturbatively, where the first order correction to the wave function from scattering for the $k^{\text{th}}$ mode is given by the following convolution of the potential with the 4-dimensional Green's function
 \ba
 \psi_{s}(\x,t) \amp=\amp \int d^4 x' G_4(t-t',\x-\x') e^{-i \Ek  t'}\nonumber\\
 \amp\amp\times \left(\A(r',-\nabla'^2)+\B(r',-\nabla'^2)\left(\frac{e^{i(2(\oma t'+\varphi))}+e^{-i(2(\oma t'+\varphi))}}{2}\right)\right)\psi^{(k)}_u(\x')
 \ea
In the first term here with $\A$, which is independent of time, we can easily carry out the integral over time and re-write the answer in terms of the 3-dimensional (spatial) Green's function. In the second term here with $\B$, which depends on time, it seems to be more complicated. However, since we are assuming a single-harmonic coherently-oscillating source, we can absorb the time dependence into the exponential $e^{-iE_kt'}$ prefactor, allowing us to once again carry out the integral over time; this time in terms of the 3-dimensional (spatial) Green's function with an appropriately shifted value of energy/frequency. This gives
\ba
&&\psi_{s}(\x,t) = \int d^3 x' \bigg\{ G_3(\Ek ,\x-\x')e^{-i\Ek t}\A(r',-\nabla'^2)\nonumber\\
&&+\frac{1}{2}\Big[G_3(E_{-1},\x-\x')e^{-i(E_{-1}t-2\varphi)}+G_3(E_{+1},\x-\x')e^{-i(E_{+1}t+2\varphi)}\Big]\B(r',-\nabla'^2)\bigg\}\psi^{(k)}_u(\x')
\,\,\,\,\,\,\,\,\,\,\,\,\,\,\,\,\,
\ea
Here the shifted energies/frequencies $E_{-1}$ and $E_{+1}$ are defined as
\beq
E_\alpha\equiv \Ek +2 \, \alpha\, \oma
\eeq
with $\alpha=\pm 1$. 
Then, as in Section~\ref{sec:wavepckt}, the unscattered wave function is expressed as a wave packet with Fourier amplitudes $\psik(\q)$. The resulting scattered wave can be written as
\ba
\psi_s(\x,t)\amp=\amp\int d^3 x' \frac{d^3q}{(2\pi)^3}\frac{ \psik(\q)}{\sqrt{2\Eq }}e^{i\q\cdot\x'}e^{-i\q\cdot\bv}\bigg\{ G_3(\Eq,\x-\x')e^{-i\Eq t}\A(r',q^2)\nonumber\\
\amp\amp+\frac{1}{2}\Big[G_3(E_{-1},\x-\x')e^{-i(E_{-1}t-2\varphi)}+G_3(E_{+1},\x-\x')e^{-i(E_{+1}t+2\varphi)}\Big]\B(r',q^2)\bigg\}\,\,\,\,\,\,
\ea
where the derivatives in the potentials $\A(r,-\nabla^2)$ and $\B(r,-\nabla^2)$ have acted on the plane waves, and now the potentials depend instead on the square of the momenta $q^2$.

Since the Green's function is most naturally written in terms of momentum and not energy, we define the following momenta
\ba
q_\alpha\amp\equiv\amp\sqrt{E_\alpha^2-m_p^2}=\sqrt{(E_q+(2\alpha \oma)^2)^2-m_p^2}
\ea
The Green's function can be approximated in the far distance regime as
\beq G_3(E_\alpha,\x-\x')=-\frac{E_\alpha}{2\pi |\x|}e^{iq_\alpha|\x|}e^{-i\q'_\alpha\cdot\x'}\eeq
where $\q'_\alpha\equiv q_\alpha\hat{x}$. 
Then the scattered wave packet becomes
\beq
\psi_s(\x,t)
=\sum_{\alpha=0,-1,+1}\int \frac{d^3q}{(2\pi)^3}e^{-i(E_\alpha t+2\alpha\varphi)}\F_\alpha(\q_\alpha',\q)\frac{e^{iq_\alpha r}}{r}\frac{\psik(\q)}{\sqrt{2\Eq }}e^{-i\q\cdot\bv}\label{psisTD}
\eeq
where we have defined a set of scattering amplitudes $\F_\alpha$ as
\ba
\F_0(\q',\q)\amp\equiv\amp -\int d^3x' \frac{E_q}{2\pi}e^{i(\q-\q')\cdot\x'}\A(r',q^2)\\
\F_{\pm1}(\q_{\pm1}',\q)\amp\equiv\amp -\frac{1}{2}\int d^3x' \frac{E_{\pm1}}{2\pi}e^{i(\q-\q_{\pm1}')\cdot\x'}\B(r',q^2)
\ea

\subsection{Decoherence Rate}
We return now to considering a DMSCS source for potential $V$, specifically a scalar field in a superposition. We will examine two cases with increasing specialization. Firstly, we will allow the centers of the potentials $V_1$ and $V_2$ to differ, as characterized by $\Lv_1,\,\Lv_2$. In addition, the sub-states of the scalar field will differ in the phase $\varphi$ as defined in eq.~(\ref{potlcos2m}), while the spatial profile of the scalar field is taken to be the same between the sub-states (up to the difference in the centers). 
Then in the next subsection, we will specialize to the case where the centers of the sub-states coincide ($\Lv_1=\Lv_2$), and thus the only difference between the potentials $V_1$ and $V_2$ will be from the phase $\varphi$ of the scalar field. This second case captures one of the goals of this paper, which is to evaluate the degree of decoherence for a configuration in which Newtonian gravity cannot detect the superposition, and thus only general relativistic effects have the ability to lead to decoherence.

As done in Section~\ref{sec:static decrate}, the calculation of the decoherence rate involves an integral of the wave function overlaps with respect to impact parameters $S_{ij}= \int d^2 b \braket{\psi_{s,i}}{\psi_{s,j}}$. Using eq.~(\ref{psisTD}) and shifting the center of mass appropriately, this is given by
\ba
 S_{ij}
\amp=\amp\int d^2 b \,d^3 x \frac{d^3 q\, d^3 \tilde{q}}{(2\pi)^6}\sum_{\alpha,\beta}\Bigg\{\left[ e^{-i \left(E_{\bt}t+2\beta\varphi_j\right)}F_\beta(\qtv_\beta',\qtv)\frac{e^{i\qt_\beta r_j}}{r}\frac{\psik(\qtv)}{\sqrt{2E_{\qt}}}e^{-i\qtv\cdot(\bv-\Lv_j)}\right]\nonumber\\
\amp\amp\times\left(E_\alpha+E_{\bt}\right)\left[ e^{+i \left(E_\alpha t+2\alpha\varphi_i\right)}F_\alpha^*(\q_\alpha',\q)\frac{e^{-iq_\alpha r_i}}{r}\frac{\psik^*(\q)}{\sqrt{2\Eq }}e^{+i\q\cdot(\bv-\Lv_i)}\right]\label{Sijbig} 
\ea
where once more we have used $r_i\equiv |\x-\Lv_i|\approx r - \hat{x}\cdot\Lv_i$. Here we have used notation that $E_{\bt}$ is defined as $E_\beta(\qt)=E_{\qt}+2\beta\oma$. We once again perform the following integrals to obtain some delta functions
\beq 
\int d^2 b\, e^{-i\bv\cdot(\q-\qtv)}=(2\pi)^2\delta^2(\q_\perp - \qtv_\perp)\label{bint}
\eeq
\beq \int d^3 x\frac{e^{-i(q_\alpha-\qt_\beta)r}}{r^2}=\int d^2\Omega\int ^\infty_0 dr \, e^{-i(q_\alpha-\qt_\beta)r}=\int d^2\Omega (2\pi) \delta(q_\alpha-\qt_\beta)\label{rint}\eeq
Using these delta functions, we can evaluate either the $d^3\qt $ integrals and replace $\qtv$ by a function of $\q$, or we can evaluate first the $d^3q$ integrals and replace $\q$ with a function of $\qtv$. In either case, the delta functions guarantee that $E_\alpha(q)=E_\beta(\qt)$ (implicit from the condition $q_\alpha=\qt_\beta$) which guarantees that the integral becomes time independent as expected.  For completeness, we should note that the delta functions provide conditions which allow $\qtv$, for example, to be replaced by more than one valid function of $\q$; however, since the distributions $\psik$ are meant to be relatively sharply peaked, we will only consider the $\qtv$ picked out by the delta functions which are approximately aligned with $\q$, since the distrubution functions will suppress the cases in which $\q$ and $\qtv$ differ greatly.

If we choose to first perform the $d^3\qt$ integral, each delta function will result in the components of $\qtv$ being replaced by functions of the components of $\q$; we denote this vector by $\qtv\rightarrow\q_{\alpha\beta}$
\beq 
q_{\alpha\beta}=\begin{cases}\sqrt{q^2+4\oma(\alpha-\beta) E_q+4\oma^2(\alpha-\beta)^2}&\,\,\,\,\,\text{for }E_\alpha\geq0\\
\sqrt{q^2+4\oma(\alpha+\beta) E_q+4\oma^2(\alpha+\beta)^2}&\,\,\,\,\,\text{for }E_\alpha<0\end{cases}
\eeq
Note that $E_\alpha\geq0$ is guaranteed if $\Eq >2\oma$. We will in general wish to examine the case where $\oma\ll \Eq ,q,m_p$; in this case we can expand to first order in $\oma$ as follows
\beq q_{\alpha\beta}=q+\frac{2 \oma}{q}E_q(\alpha-\beta)+\mathcal{O}(\oma^2)\eeq
In addition, the cancelation of the time dependence following the $d^3\qt$ integration results in a different leftover phase for each term, given simply by $2(\alpha\varphi_i-\beta\varphi_j)$. Then, the overlap integral can be written more concisely as
\begin{align} 
S_{ij}=\sum_{\alpha,\beta}\int d^2\Omega\int& \frac{d^3q}{(2\pi)^3}\bigg\{F^*_\alpha(\q_\alpha',\q)F_\beta(\q'_{\alpha},\q_{\alpha\beta})\nonumber\\
&\times e^{-i(\q-\q'_\alpha)\cdot\Lv_i}e^{+i(\q_{\alpha\beta}-\q'_{\alpha})\cdot\Lv_j}e^{2i(\alpha\varphi_i-\beta\varphi_j)}\psik^*(\q)\psik(\q_{\alpha\beta})\bigg\}\label{SijL}
\end{align}
where in the last expression, we have simplified using $\frac{E_\alpha}{\sqrt{\Eq  E_{q_{\alpha\beta}}}}\approx 1$, which is valid for $\oma$ which is small compared to $\Eq$. This is required for self-consistency, since this is the regime in which the source is unable to pair produce particles, and we are indeed working in the single particle formalism. For completeness, the full expression with this factor is provided in Appendix~\ref{app:sab}, but it will not be needed for the remainder of our analysis.

We note that when computing $S_{ij}$, the only dependence on $i$ or $j$, i.e., whether the potential is $V_1$ or $V_2$, is in the form of phases and the position of the center of mass (by construction). The result of eq.~(\ref{SijL}) can be used to compute the decoherence rate for this configuration.

\subsection{Phase Difference}\label{sec:phases}

We will specialize now to a case where the superposition is one in which the  spatial mass distribution is entirely the same, including the location of the centers of mass ($\Lv_1=\Lv_2=0$), and only the phase $\varphi$ of the axion field differs. Thus, the only part of the integrals which depends on the indices $i$ and $j$ is the leftover phase $e^{2i(\alpha\varphi_i-\beta\varphi_j)}$ which can be taken out of the integral. We can thus write the integrals $S_{ij}$ as
\beq S_{ij}=\sum_{\alpha,\beta}s_{\alpha\beta}e^{2i(\alpha\varphi_i-\beta\varphi_j)}\label{Ssum}\eeq
where the coefficients $s_{\alpha\beta}$ are defined as
\beq 
s_{\alpha\beta}\equiv\int d^2\Omega\int \frac{d^3q}{(2\pi)^3}\bigg\{\F^*_\alpha(\q_\alpha',\q)\F_\beta(\q'_{\alpha},\q_{\alpha\beta})
\psik^*(\q)\psik(\q_{\alpha\beta})\bigg\}\label{sab}
\eeq

Then, to compute the decoherence rate, we consider the combination of $S_{ij}$ as required in eq.~(\ref{Delta}) to compute $\Delta$.
It can now be seen that any part of $S_{ij}$ which does not depend on the phase $\varphi$ will simply cancel when computing $\Delta$. Explicitly, this is
\ba 
S_{11}+S_{22}-2\Re\left[S_{12}\right]
\amp=\amp 2(1-\cos(2(\varphi_1-\varphi_2)))(s_{-1,-1}+s_{+1,+1})\nonumber\\
\amp\amp+2\left\{\cos(4\varphi_1)+\cos(4\varphi_2)-2\cos(2(\varphi_1+\varphi_2))\right\}\Re[s_{-1,+1}]
\label{s1122}
\ea
where we have used the fact that $\Re[s_{\alpha,\beta}]=\Re[s_{\beta,\alpha}]$ to simplify (see Appendix~\ref{app:sab}). Not only does the term $s_{0,0}$, completely independent of the potential's index $i$ or $j$, vanish, but also the terms of the form $s_{0,\beta}$ or $s_{\alpha,0}$. Thus, in this case, one must only compute $s_{-1,-1}$, $s_{+1,+1}$, and $\Re[s_{-1,+1}]$.

Then, after carrying out the computation of the necessary $s_{\alpha,\beta}$, the decoherence rate is
\ba 
\Gamma_{dec}
\amp=\amp\frac{1}{2}n v\bigg\{2(1-\cos(2(\varphi_1-\varphi_2)))(s_{-1,-1}+s_{+1,+1})\nonumber\\
\amp\amp+2\left\{\cos(4\varphi_1)+\cos(4\varphi_2)-2\cos(2(\varphi_1+\varphi_2))\right\}\Re[s_{-1,+1}]\bigg\}
\ea
where $n$ is the local number density for the probe particles and $v$ is their typical velocity.

Another interesting feature of this scenario arises when considering the momentum distribution functions $\psik(\q)$. One typically expects that the probe particle has well-enough-defined momentum such that the dynamics are well approximated by considering a plane wave (a wave packet where $\psik(\q)\propto \delta(\q-\k)$). For narrowly peaked $\psik$, one expects the product $\psik^*(\q)\psik(\q_{\alpha\beta})$ to be exponentially supressed unless $\alpha=\beta$. Specifically, two regimes of interest are when
\begin{enumerate}[(A)]
\item The width of the $\psik$ distribution is large compared to the difference between $\q$ and $\q_{\alpha\beta}$ (roughly $\sim\oma(\alpha-\beta)E_q/q$).
\item The width of the distribution is small compared to this difference.
\end{enumerate}
In case (A), the product $\psik^*(\q)\psik(\q_{\alpha\beta})$ is well approximated by $|\psik(\q_{av})|^2$, where $\q_{av}$ is some intermediate value between $\q$ and $\q_{\alpha\beta}$. In case (B), however, the product $\psik^*(\q)\psik(\q_{\alpha\beta})$ will necessarily cause the integral to be exponentially suppressed, except for when $\alpha=\beta$ which gives $|\psik(\q)|^2$. The condition for (A) is satisfied when the spatial variance of the probe particle wave packet is smaller than the spatial variance of the DMSCS, thus we expect case (A) to be of more physical relevance.

Both cases (A) and (B) are of interest since they can both lead to decoherence. In addition, both of these regimes allow for the simplification of eq.~(\ref{sab}). For situation (A), we can approximate the distributions as being evaluated at $\q_{av}$ as $|\psik(\q)|^2$, while in situation (B) the momentum distributions only contribute to the integral when $\alpha=\beta$ in the form $|\psik(\q)|^2$. For either case, let us presume that the composite distribution $|\psik(\q_*)|^2$ (where $\q_*$ is either $\q_{av}$ in situation (A) or $\q$ in (B)) is sufficiently narrow such that the integral in eq.~(\ref{sab}) is well approximated by evaluating the rest of the integrand at $\q_*=\k$, giving
\beq
s_{\alpha\beta}
\approx \sig_{\alpha\beta}(\q)\Big|_{\q_*=\k}\label{Sigma}
\eeq
where
\beq
\sig_{\alpha\beta}(\q)\equiv \int d^2\Omega\,\F^*_\alpha(\q_\alpha',\q)\F_\beta(\q'_{\alpha},\q_{\alpha\beta})\label{Sigmadef}
\eeq
is another generalized cross section. 
%When the oscillation frequency is small compared to the energy of the probe particle, the fraction in eq.~(\ref{Sigma}) is approximately 1, such that 
Then, the decoherence rate can be more simply expressed in terms of the generalized cross section as
\ba
\Gamd\amp\approx\amp\frac{1}{2}n v\bigg\{2(1-\cos(2(\varphi_1-\varphi_2)))(\sig_{-1,-1}+\sig_{+1,+1})\nonumber\\
\amp\amp+2\left\{\cos(4\varphi_1)+\cos(4\varphi_2)-2\cos(2(\varphi_1+\varphi_2))\right\}\Re[\sig_{-1,+1}]\bigg\}\label{GammaSigma}
\ea
or, in the case (B) above where $\sig_{-1,+1}$ is exponentially suppressed, the decoherence rate is more simply
\beq
\Gamd\approx\frac{1}{2}n v\bigg\{2(1-\cos(2(\varphi_1-\varphi_2)))(\sig_{-1,-1}+\sig_{+1,+1})\bigg\}\label{GammaSigma2}
\eeq

\subsection{Parameterization and Gaussian Example}\label{sec:gausstd}

We seek once more to parameterize the decoherence rate in such a way that makes the scales in the problem manifest. In the interest of clarity, we will consider an example configuration for the scalar field.
We study a Gaussian spatial profile for the scalar field 
\beq
\phi(r,t)=\sqrt{2}\,\phia\, e^{-r^2\mu^2/2}\cos(\omega t +\varphi)
\eeq
with $\phi_a$ an amplitude and $\mu$ an inverse length scale. 
Upon substitution into eq.~(\ref{deltaEqn}) we can carry out the integrals to solve for $\delta$, and obtain
\beq
\delta=-4\pi G_N \phia^2\,e^{-r^2\mu^2}\cos^2(\omega t+\varphi)
\eeq
In this form, the energy density of eq.~(\ref{rhocos2}) becomes
\beq 
\rho = \phia^2e^{- r^2\mu^2}\left[\left(\frac{\ma^2 - \oma^2+ r^2\mu^4 }{2}+\oma^2 \right)+\left(\frac{ \ma^2 - \oma^2+ r^2\mu^4 }{2}\cos(2(\oma t+\varphi))\right)\right]
\eeq

Introducing now the mass scale $M$ associated with the energy density of the configuration, we wish to write the energy density $\rho$ in terms of a dimensionless function $\zeto$ (analogous to the function $\zeta$ we introduced in the static case) of the dimensionless variables $\hat{x}\equiv \mu r$ and $\hat{T}\equiv \omega t$
\beq 
\rho = M \mu^3 \zeto(\hat{x}, \hat{T})
\eeq
To do this, we write 
\beq
\phia^2 = {\kappa M\mu^3\over m_a^2}
\eeq
 where $\kappa$ is an $\mathcal{O}(1)$ dimensionless constant and $\zeto(\hat{x}, \hat{T})$ is identified as
\beq 
\zeto(\hat{x}, \hat{T}) = \frac{\kappa}{m_a^2}\frac{e^{-\hat{x}^2}}{2}\left[\left(\ma^2+ \oma^2+ \hat{x}^2\mu^2 \right)+\left(\ma^2 - \oma^2+ \hat{x}^2\mu^2 \right)\cos(2(\hat{T}+\varphi))\right]
\eeq

Since the scattering amplitudes are related to Fourier transforms of the potentials with respect to some transfer momentum $\pv_{tr}\equiv\q-\q'_\alpha$, we can write first the Fourier transforms of the potentials according to the definition in eq.~(\ref{fourierdef}) (which transforms with respect to $\hat{\pv}\equiv\pv_{tr}/\mu)$. Taking the Fourier transform gives a form analogous to eq.~(\ref{PoissonFT})
\beq 
-\hat{p}^2\,\hat{\Psi}(\hat{p},\hat{T})= 4\pi G_N M \mu\, \hat{\zeto}(\hat{p},\hat{T})
\eeq
We can also write the form of the potential $\Phi(r,t)$ in terms of the dimensionless variables, using $\Phi(r,t) = \Psi(r,t)+\delta(r,t)$.
To consolidate notation, we define the following two dimensionless functions of $\hat{x}$ and their Fourier transforms as functions of $\hat{p}$
\ba 
&&g(\hat{x})\equiv e^{-\hat{x}^2},\,\,\,\,\,\,\,\,\,\,\,\,\,\,\hat{g}(\hat{p})= \pi^{3/2} e^{-\frac{\hat{p}^2}{4}}\\
&&h(\hat{x})\equiv \hat{x}^2 \, e^{-\hat{x}^2},\,\,\,\,\,\,\, \hat{h}(\hat{p})=-\frac{\pi^{3/2}}{4}e^{-\frac{\hat{p}^2}{4}}(\hat{p}^2-6)
\ea
such that $\zeto$ becomes
\beq \zeto(\hat{x},\hat{T})=\frac{\kappa\mu^2}{2m_a^2}\left[\frac{\ma^2+\oma^2}{\mu^2}g(\hat{x})+h(\hat{x})+\left(\frac{\ma^2-\oma^2}{\mu^2}g(\hat{x})+h(\hat{x})\right)\cos(2(\hat{T}+\varphi)\right]
\eeq
and the Fourier transforms of the potentials become
\ba
\hat{\Psi}(\hat{p},\hat{T})\amp=\amp2\pi G_N {\kappa M \mu^3\over m_a^2} \hat{g}(\hat{p})\!\left[\left(\frac{1}{4}-\frac{\ma^2+\oma^2}{\mu^2\hat{p}^2}-\frac{3}{2\hat{p}^2}\right)+\left(\frac{1}{4}-\frac{\ma^2-\oma^2}{\mu^2\hat{p}^2}-\frac{3}{2\hat{p}^2}\right)\cos(2(\hat{T}+\varphi))\right]\\
\hat{\Phi}(\hat{p},\hat{T})\amp=\amp-2\pi G_N {\kappa M \mu^3\over m_a^2} \hat{g}(\hat{p})\!\left[\left(\frac{1}{4}+\frac{\ma^2+\oma^2}{\mu^2\hat{p}^2}+\frac{3}{2\hat{p}^2}\right)+\left(\frac{1}{4}+\frac{\ma^2-\oma^2}{\mu^2\hat{p}^2}+\frac{3}{2\hat{p}^2}\right)\cos(2(\hat{T}+\varphi))\right]\,\,\,\,\,\,\,\,\,\,\,\,\color{white}a
\ea

We can now readily separate the Fourier transforms of the potentials into the time dependent and independent parts
\ba \Ahat(\hat{p},q^2)\amp=\amp2\pi G_N {\kappa M \mu^3\over m_a^2} \hat{g}(\hat{p})\left[-\left(\frac{1}{4}+\frac{\ma^2+\oma^2}{\mu^2\hat{p}^2}+\frac{3}{2\hat{p}^2}\right)E_q+\left(\frac{1}{4}-\frac{\ma^2+\oma^2}{\mu^2\hat{p}^2}-\frac{3}{2\hat{p}^2}\right)\frac{q^2}{E_q}\right]{\color{white}aaa}\\
\Bhat(\hat{p},q^2)\amp=\amp2\pi G_N {\kappa M \mu^3\over m_a^2} \hat{g}(\hat{p})\left[-\left(\frac{1}{4}+\frac{\ma^2-\oma^2}{\mu^2\hat{p}^2}+\frac{3}{2\hat{p}^2}\right)E_q+\left(\frac{1}{4}-\frac{\ma^2-\oma^2}{\mu^2\hat{p}^2}-\frac{3}{2\hat{p}^2}\right)\frac{q^2}{E_q}\right]\color{white}aaa\ea
Finally, we can write the scattering amplitudes as
\ba 
\F_0(\q',\q) \amp=\amp -\frac{E_q}{2\pi\mu^3}\Ahat(\hat{p},q^2)\bigg|_{\hat{p}=\frac{\q-\q'}{\mu}}\label{F0gauss}\\
\F_{\pm1}(\q_{\pm1}',\q)\amp=\amp -\frac{1}{2}\frac{E_{\pm 1}}{2\pi\mu^3}\Bhat(\hat{p},q^2)\bigg|_{\hat{p}=\frac{\q-\q'_{\pm1}}{\mu}}\label{F1gauss}
\ea

In $F_0$, the exponential $\hat{g}$ in $\Ahat$ and $\Bhat$ is $\cO(1)$ when the transfer momentum goes to zero ($\theta=0$, the forward direction), but this is also where these functions suffer from a divergence. This is the well-known divergence that comes from an infinite-range potential (e.g. the coulomb potential and the forward divergence of the Rutherford scattering amplitude). It arises because the density $\rho$ which sources the potential has a monopole. In this case one can still proceed by restricting to configurations that have a vanishing monopole; this is a physical thing to do, because the monopole will decohere quickly, leaving other pieces of the distribution in a quantum state. 
Importantly, however, in the case in which the 2 parts of DMSCS differ only in phase, any divergences from $F_0$ will exactly cancel, and thus it is possible to calculate the decoherence rate for that configuration using this scattering formalism. We will focus on this case later in the paper.

For $F_{\pm1}$, when $\alpha\neq0$, the transfer momentum cannot be exactly zero because $\q$ and $\q_\alpha'$ have different magnitudes. Thus, there is in principle no forward divergence. But now the exponentials in the amplitudes are no longer guaranteed to be $\cO(1)$, and thus the amplitude may become exponentially suppressed. 

Let us examine one regime of interest, where the oscillation frequency is small compared to the energy scales of the probe particle. Specifically, we take $\frac{4\oma E_q}{q^2}$, $\frac{4\oma^2}{q^2}\ll1$, and thus we expand the (dimensionless) transfer momentum to second order in $\oma$ as follows
\ba
\hat{p}^2\amp=\amp \frac{(\q-\q'_{\alpha})^2}{\mu^2}
\approx\frac{1}{\mu^2}\left(\left(2q^2+4\alpha \oma E_q\right)(1-\cos{\theta})+\frac{4\alpha^2\oma^2\left(q^2+(E_q^2-q^2)\cos{\theta}\right)}{q^2}\right)
\ea
In the forwards direction $\theta\to0$ this becomes
\ba
\hat{p}^2\bigg{|}_{\theta=0}
\amp\approx\amp \frac{4\alpha^2\oma^2 E_q^2}{\mu^2 q^2}\label{phatalpha}
\ea
The forwards direction is important since this is where the transfer momentum is at a minimum, and we wish to examine how far from zero the transfer momentum must be. The leftover expression can be interpreted in the following way: For a probe which is not relativistic, $E/q\sim m_p/q\sim 1/v_p$. In addition, the length of the region containing the DMSCS is given approximately by $L\sim1/\mu$, and so the product $\frac{E}{\mu q}\sim {L/v_p}\sim t_c$ where $t_c$ is the time it takes the probe to cross through this region. Then, $\frac{\oma E}{\mu q}\sim \oma t_c$ is approximately the number of oscillations of the DMSCS the probe particles encounters during its scattering. Estimating the number of oscillations as 
\beq
N\equiv\frac{\oma\, E}{\mu \,q}
\eeq
we have that the minimum transfer is $\hat{p}^2\big{|}_{\theta=0}\approx 4\alpha^2 N^2$. 

We first note that when $\alpha=0$, corresponding to the amplitude $F_0$, then the limit $\theta\to0$ gives $e^{-\hat{p}^2/4}\to1$ (as indicated previously, along with the forward divergence). However, with nonzero $\alpha$, depending on parameters, it may be the case that the scattering amplitude, and therefore scattered wave function, is itself suppressed, and exponentially so. That is, for $N\gg1$, the rapid oscillations of the scalar field are effectively ignored by the probe. Thus, any contribution to decoherence from the $\alpha\neq0$ part of the scattered wave function would be miniscule. In this regime, decoherence would proceed in approximately the same way as in the static case, effectively ignoring the time-dependence of the potential.

\section{Application to Dark Matter in Galaxy}\label{sec:disc}

In this section, we apply these results to a dark-matter-Schr\"{o}dinger-cat-state (DMSCS) in the galaxy today. Our results apply to any light scalar (or light boson) as the DM, with a primary motivations from axions. The sub-states of the superposition of the DMSCS will differ only in the phase of the axion field. If we instead consider a DMSCS with sub-states that differ in the spatial profile of the mass-distribution, the decoherence rate will be dominated by this spatial difference, resulting in a decoherence rate more simply found using the static and nonrelativistic analyses. Thus we restrict to the case of only differing phases.

We will continue to use the results from the Gaussian profile discussed in the previous section as a prototype for a DMSCS. The phenomena resulting from this Gaussian analysis should generalize fairly generically. For example, we will see in the next section that in the nonrelativistic limit, the effects of scattering which correspond to the oscillation of the scalar field will be suppressed because they reside in the tail of the momentum distribution of the Gaussian profile. This momentum distribution, also a Gaussian with a width of the order of the characteristic momentum, is similar to the Maxwell velocity distribution that one expects to find for DM in the galaxy.

\subsection{Nonrelativistic  Dark Matter}\label{sec:nonrelaxion}
For ordinary models of DM in the galaxy, one expects diffuse DM particles to have velocity dispersion to be far lower than that of light. The reason is that diffuse DM tends to virialize and we know that the virial speed of matter in the galaxy is small. Furthermore, if the diffuse DM were moving relativistically, it would easily escape the galaxy. The discussion in the previous section suggests that this makes the decoherence of phases to be exponentially suppressed; however, we shall quantify this more precisely here.

For diffuse nonrelativistic DM, such as axions, we can take the oscillation frequency to be approximately the mass of the DM, $\oma\approx\ma$. In addition, the scale $\mu$ is set by the de Broglie wavelength of the DM such that $\mu\sim p_a =m_a v_a$. We can consider a nonrelativistic particle as a probe, such as a baryon/proton in the galaxy. Then using the typical DM and baryon speeds of the order of the virial speed in our galaxy $\sim 10^{-3}$ (in units where $c=1$) \cite{Kuhlen:2012ft}, we obtain a minimum transfer momentum of
\beq 
-\frac{\hat{p}^2}{4}\bigg{|}_{\theta=0} \approx -\frac{\oma^2 (m_p^2 + k_p^2)}{m_a^2 v_a^2 k_p^2}\approx
-\frac{1}{v_a^2 v_p^2}\sim -10^{12}
\eeq
Similarly, for a photon as the probe, or relativistic cosmic ray protons, we have a minimum transfer momentum of
\beq 
-\frac{\hat{p}^2}{4}\bigg{|}_{\theta=0} \approx -\frac{\oma^2 }{ m_a^2 v_a^2}\approx-\frac{1}{ v_a^2}\sim - 10^6
\eeq

Thus, since these quantities appear in the arguments of exponentials in the scattering amplitudes, we see that the decoherence rate of the phase is highly exponentially suppressed. This suppression in scattering seems compatible with the time dependent 1-dimensional problem studied in Ref.~\cite{Byrd2012}.
We therefore only expect an appreciable response from scattering corresponding to when $\hat{p}$ truly vanishes. This corresponds to the generalized cross section $\sig_{0,0}$, which means that the only appreciable contribution to decoherence comes from the part of the wave function that is independent of the phases of the scalar field. Therefore, if the superposition involves spatially distinct profiles, the decoherence rate is well described by the decoherence rate for the static metric in Section~\ref{sec:static decrate}, and can be appreciable. If instead the superposition is only of phases as in Section~\ref{sec:phases}, one finds a decoherence rate that is exponentially suppressed. Such a configuration is therefore quite robust against decoherence; we shall return to a discussion of possible implications of this for earth based searches for axions in Section \ref{Exp}.

\subsection{Relativistic Dark Matter}\label{sec:relaxion}

Considering again axions as the DM, let us now consider a piece of the DM that involves (semi)-relativistic axions. This idea may seem at odds with the idea of cold-dark-matter (CDM), which is known to fit the data well. So let us clarify what we mean by this: (i) we can consider just a small fraction of the DM to be relativistic, such as that which is near black holes (as we will discuss in Section \ref{BH}), or (ii) we can even be considering the bulk of the DM particles to be relativistic, so long as they are in bound clumps, like boson stars, whose center-of-mass is moving slowly so as to act as CDM. But the velocity {\em dispersion} around the center of mass within a clump could be relativistic (as we will discuss in Sections \ref{BS} and \ref{Osc}).

In this situation, we have $\mu\sim p_a \gtrsim m_a$. It is no longer guaranteed that the exponentials in the scattering amplitudes lead to an exponentially suppressed contribution to the wave function from scattering. Furthermore, one does also need relativistic probes, such as cosmic ray protons or photons, which we shall discuss in the next section.

To study the regime in which the exponentials of the scattering amplitudes are appreciable, we consider the case where both the probe and the axion have relativistic energies. One should still expect the energy of the axion to be much smaller than that of the probe, so we continue to assume this relationship (this can be satisfied trivially, for example with a relativistic proton or photon probe). Then, indeed, the transfer momentum as expressed in eq.~(\ref{phatalpha}) is $\cO(1)$ and thus the exponential factors in the scattering amplitudes will not suppress the amplitudes dramatically.

In this regime, we can make use of either eq.~(\ref{GammaSigma}) or eq.~(\ref{GammaSigma2}) to estimate the decoherence rate, where the choice between these two expressions depends on the details of the wave packet for the probe particle. As discussed in Section~\ref{sec:phases}, if the width of the distribution function $\psik(\q)$ is large compared to the difference between $\q$ and $\q_{\alpha\beta}$ (case (A)), then eq.~(\ref{GammaSigma}) is the appropriate approximation for the decoherence rate. Instead, if the width of the distribution is small compared to this difference (case (B)), then eq.~(\ref{GammaSigma2}) is more appropriate. In either case, we will simply evaluate the expressions at $\q = \k$ (though in case (A), in principle, $\k$ should be set to some $\q_{av}$, this approximation will suffice to estimate the decoherence rate). To evaluate these expressions, one must compute the generalized cross sections defined in eq.~(\ref{Sigmadef}) explicitly.
The form of these cross sections is quite complicated, but can be simplified greatly to achieve an order-of-magnitude estimate for the decoherence rate; the details of this approximation and the resulting cross sections are discussed in Appendix~\ref{app:decrate estimate}. Here, we give the resulting decoherence rate in terms of the physical parameters defined in Section~\ref{sec:gausstd}
\beq
\Gamd\approx G_N^2 (\kappa M)^2 n  v \, e^{-\frac{2\oma^2 E_k^2}{k^2 \mu^2}}\frac{\pi\left(2 k^4 (3\mu^2-\oma^2)+3k^2\mu^2m_p^2+\oma^2m_p^4+k^2\ma^2(2k^2+m_p^2)\right)^2}{4 k^2 \mu^2 \oma^4 (k^2 +m_p^2)^2} \phis(\varphi_1,\varphi_2)
\label{Gamrel}\eeq
where $\phis$ is an $\mathcal{O}(1)$ function of the phases; its limiting values are
\ba
\phis_A(\varphi_1,\varphi_2)\amp=\amp \frac{\left(2 + \cos{4\varphi_1}+\cos{4\varphi_2}-4\cos(2(\varphi_1-\varphi_2))\right)}{8}\\
\phis_B(\varphi_1,\varphi_2) \amp=\amp \frac{\sin^2(\varphi_1-\varphi_2)}{2}
\ea
where the choice depends on whether one is in case (A) with eq.~(\ref{GammaSigma}) ($\phis_A$) or case (B) with eq.~(\ref{GammaSigma2}) ($\phis_B$). 

We can set the mass scale $M$ to be the amount of DM mass contained in a region of volume $1/\mu^3$ by specifying the local DM density $\rho_{\DM}$; these are roughly related as $\rho\approx m_a^2\phi_a^2=\kappa\, M\,\mu^3$. In addition, for relativistic particles, we can relate their momenta and their energies/masses to further eliminate parameters. For a massive probe, we parameterize the momentum of the probe as $a\equiv k/m_p$ where $\an$ is a numerical factor which can be $\cO(1)$. In this case, we can replace the mass of the probe particle in favor of its energy by $m_p= E (1+a^2)^{-1/2}$ where $E=E_k$ is the energy of the probe, and the speed of the probe is simply $v = a(1+a^2)^{-1/2}$. We can also eliminate the number density of probes in favor of their energy density as $n=\rho_p/E$. If instead the probe is massless, then $v=1$, $m_p=0$, and $k=E$. 
Similarly, for a semi-relativistic DM, we can take the oscillation frequency to be on the order of the mass of the axion; we parameterize it by $\bn\equiv\oma/\ma$. Also, for semi-relativistic DM the inverse length scale $\mu$ can also be on the order of the axion mass; we parameterize it by $\cn\equiv\mu/\ma$. 
With this parameterization, we obtain the following result for the decoherence rate
\beq \Gamd\approx \abc \frac{G_N^2 E\, \rho_p\, \rho_{\DM}^2}{\ma^8}\,\phis(\varphi_1,\varphi_2)\label{Gammasimple}\eeq
where $\abc$ is a dimensionless numerical factor, which includes the exponential factor. For a massive probe particle it is given by
\beq 
\abc_{m_p} \equiv e^{-\frac{2(1+\an^2)\bn^2}{\an^2\cn^2}} \frac{\pi\left(\bn^2+\an^2(1+3\cn^2)+\an^4(2-2\bn^2+6\cn^2)\right)^2}{4\an(1+\an^2)^{7/2}\bn^4\cn^8}
\eeq
For a massless probe particle ($a\to\infty$) this simplifies to
\beq 
\abc_0\equiv e^{-\frac{2\bn^2}{\cn^2}} \frac{\pi(1-\bn^2+3\cn^2)^2}{\bn^4\cn^8}
\eeq

The decoherence rate can then be obtained by specifying the densities and energies of the probe particle and the axion. As a starting point, we can compare density estimates for the DM to the local average density of DM in the Milky Way \cite{Read:2014qva}
\beq
\rho_{loc,mw} \approx 0.4 \frac{\mbox{GeV}}{\mbox{cm}^3}
\eeq
Although, as we discuss below, what really matters is the energy density of {\em relativistic} DM (which may be DM near black holes, or DM that is in the form of dense clumps, etc), so this figure is not directly relevant, but acts as merely a starting point for consideration. 

We can consider different types of probe particles. Most of the baryonic mass in the galaxy is made of hydrogen atoms; for simplicity, we can think of these as being protons. Since we are considering only (semi)-relativistic probes, we can take the proton to have a kinetic energy on the order of $\mbox{GeV}$. The average density in the universe is about one-fifth that of DM. So the average density at our radius from the center of the galaxy for baryons is about $\rho_b\approx\rho_{loc,mw}\approx0.08\,\mbox{GeV}/\mbox{cm}^3$. This number is evidently larger in the disk of the galaxy and much larger in the solar system, etc. However, almost all of the baryons are nonrelativistic, leading to a huge exponential suppression in the decoherence of the phase, as explained above. So our interest is in the energy density of protons that are relativistic (or at least semi-relativistic). At a typical point in the galaxy, we can use estimates of the cosmic ray density of protons, which has been estimated to be  \cite{Persic:2017qxo}
\beq
\rho_{cr}\sim 10^{-9}{\mbox{GeV}\over\mbox{cm}^3}
\eeq
So it is suppressed from the average density of nonrelativistic baryons by about 7 orders of magnitude, or so. It can be altered in the solar system due to solar or earth magnetic fields, but we shall not go into those details here.

For this baryonic probe, we calculate the following reference decoherence rate in terms of mass of the axion $\ma$, the density of DM $\rho_{\DM}$, the density of protons $\rho_p$, and the factor $\abc$. We take an example value for $\abc_{m_p}$ as $10^{-4}$ and a reference mass $\ma$ corresponding to a decoherence rate roughly near the Hubble rate today ($H_0\approx2.2\times10^{-18}\,\mbox{s}^{-1}$)
\beq 
\Gamd \sim 10^{-21}\,\mbox{s}^{-1}\!\left(\frac{\abc}{10^{-4}}\right) \left(\frac{10^{-12} \mbox{eV}}{\ma}\right)^{\!8}\left(\frac{E}{2\,\mbox{GeV}}\right)\left(\frac{\rho_{\DM}}{0.4 \,\mbox{GeV}/\mbox{cm}^3}\right)^{\!2}\left(\frac{\rho_p}{10^{-9}\,\mbox{GeV}/\mbox{cm}^3}\right) 
\eeq
This suggests the decoherence rate is quite small, unless the axion mass is somewhat below $10^{-12}$\,eV. But this conclusion is still subject to the choice of density $\rho_{\DM}$ of the DMSCS. For diffuse DM, it normally has only a very small amount that is relativistic. But under some circumstances, this density can be large; we shall return to this in the next Section.

Additionally, photons from the CMB radiation are numerous and may have a chance to decohere the source. An example value for $\abc_0$ is $10^{-2}$. The number density of CMB photons today is known to be approximately $400/\mbox{cm}^3$. We take the energy of a typical CMB photon to be $6\times10^{-4}$\,eV, which gives the following estimate for the decoherence rate 
\beq 
\Gamd \sim 10^{-16}\,\mbox{s}^{-1}\!\left(\frac{\abc}{10^{-2}}\right) \left(\frac{10^{-14}\mbox{eV}}{\ma}\right)^{\!8} \left(\frac{E}{6\times10^{-4}\,\mbox{eV}}\right)^{\!2}\left(\frac{\rho_{\DM}}{0.4 \,\mbox{GeV}/\mbox{cm}^3}\right)^{\!2}\left(\frac{n_p}{400 /\mbox{cm}^3}\right)
\eeq
Note that the approximations necessary for this decoherence rate estimate require that the mass of the axion is larger than the energy of the photon. Thus, this result should only be trusted for axion masses smaller than $10^{-4}$\,eV (or the photon energy of interest).
We see that indeed this decoherence rate is comparable to the current Hubble rate, or faster, for $m_a\lesssim 10^{-14}$\,eV. These are very light for QCD axions, but may be a type of stringy motivated axion or other ultralight axion. But again, any such conclusions are very sensitive to the local density $\rho_{\DM}$ of the DMSCS, which we discuss shortly.

\section{Discussion and Various Applications}\label{sec:outlook}

In this discussion section, we analyze our results in various contexts. Since the decoherence rate of the phase is exponentially suppressed for slowly moving DM particles, we begin by discussing contexts in which it is relativistic; including spreading in Section~\ref{Spread}, dense boson stars in Section~\ref{BS}, dense oscillons in Section~\ref{Osc}, and near black holes in Section~\ref{BH}. On the other hand, in earth based experiments, the DM should be nonrelativistic, leading to negligible decoherence of the phase, which we then discuss in Section~\ref{Exp}. We also discuss other interactions in Section~\ref{OtherInt}.

\subsection{Spatial Spreading of State }\label{Spread}

To estimate the characteristic amount of time it takes for decoherence to occur, it is usually sufficient to take the inverse of the above decoherence rate ($t_{dec}\approx \Gamd^{-1}$). However, since one may consider the DMSCS to be a fluctuation in the DM density with a scale set by its de Broglie wavelength, one should also consider that this fluctuation may spread out over time. If the characteristic time of spreading is large compared to the inverse of the decoherence rate, then the inverse of the decoherence rate is an accurate estimate for the decoherence time. Otherwise, it is necessary to compute the decoherence time as follows.

Since the DM is modeled here as a scalar (axion), it evolves under the Klein-Gordon equation. Since the DMSCS is taken to be a Gaussian-shaped fluctuation about the mean of the DM density, we can track the spatial spreading by considering the evolution of a Gaussian wave packet under the Klein-Gordon equation (as discussed, for example, in Ref.~\cite{Naumov:2013uia}). Evolving such a wave packet in time results in the growth of its spatial variance $\Delta x$; in the case of the Gaussian profile for the axion discussed in the previous sections, this is equivalent to the shrinking of the parameter $\mu$ as
\beq \Delta x \sim {1\over\mu(t)} = {1\over\mu_0}\sqrt{1+\left(\frac{t}{\tau}\right)^2}
\label{muoft}
\eeq
where the characteristic time $\tau$ is given by
\beq
\tau\equiv{\gamma \ma\over\mu_0^2}
\label{tau}
\eeq
and where $\ma$ is the scalar's mass, $\mu_0$ is the initial $\mu$, and $\gamma$ is the usual relativistic Lorentz factor for the center of mass of the Gaussian. Since will are studying the DMSCS in its rest frame, we can set $\gamma=1$ here.
Note that this shrinking does not apply directly to the momentum distribution for the particles (as momentum is conserved for free particles), but it applies to the momentum distribution for the {\em density} and hence gravitational potential; this is because they are associated with the {\em square} of the field and are affected by the spatial spread. Note that at late times, the width is growing as $\Delta x\sim 1/\mu\approx(\mu_0/m_a)t$, and by noting that one anticipates $\mu_0\sim m_a v$ for some characteristic speed $v$, this gives $\Delta x\sim v\,t$ as expected.

Then, one can write the decoherence rate as a function of time $\Gamd(t)$ by inserting eq.~(\ref{muoft}) into eq.~(\ref{Gamrel}). We then define the decoherence time $\tid$ as the solution to the following equation
\beq \int^{\tid}_0 \Gamd(t) dt = 1 \label{tidint}\eeq
For $t\ll\tau$, $\Gamd(t)\approx\Gamd(0)$, and therefore, we would have simply that $\tid=\Gamd^{-1}(0)$ so long as this $\tid$ is found to be in the regime $t\ll\tau$. On the hand, when $t\gg\tau$, we have $\mu(t)\ll\mu_0$. The decoherence rate can be estimated in either regime as
\beq \Gamd(t)\approx \Gamd(0)\left(\frac{\mu_0}{\mu(t)}\right)^2 e^{-\frac{2\oma^2 E_k^2}{k^2}\left({1\over\mu(t)^2}-{1\over\mu_0^2}\right)}
\eeq
In this late-time limit, there is not likely to be a solution to eq.~(\ref{tidint}). The decrease in $\mu$ at late times would lead to an exponentially slow rate of decoherence. Therefore, if the decoherence does not occur within a time that is of the order of $\tau$, then one should not expect decoherence to occur at all. One expects a larger $\tau$ for nonrelativistic axions. Thus it may be interesting to consider conditions that cause a nonrelativistic DMSCS, which is robust against decoherence and has slower spreading, to become relativistic and experience decoherence before sufficient spreading has occurred.

Additionally, the $\tau$ of eq.~(\ref{tau}) is only for a Gaussian evolving with the free theory Klein-Gordon equation. If, instead, one considers a DMSCS which experiences some binding force, then the spreading of the Gaussian would be slowed or stopped completely. The realization of such conditions is discussed in the following sections. 

\subsection{Boson Stars}\label{BS}

A superposition of phases of the axion (or similar light scalar DM), as discussed in Section~\ref{sec:phases}, may be of unique observational interest in the case of boson stars. Boson stars are gravitationally bound Bose-Einstein condensates of scalars. Much work has been done regarding the properties of boson stars, specifically in the contexts of axions as DM; see Refs.~\cite{ Tkachev:1986tr,Gleiser:1988rq,Seidel:1990jh,Tkachev:1991ka,Jetzer:1991jr,Kolb:1993zz,Liddle:1993ha,Sharma:2008sc,Chavanis:2011zi,Chavanis:2011zm,Liebling:2012fv,Visinelli:2017ooc,Schiappacasse:2017ham,Hertzberg:2018lmt,Hertzberg:2018zte,Levkov:2018kau,Hertzberg:2020dbk}. (Note that the condensate is a short-ranged localized clump, due to the attractive nature of gravity \cite{Guth:2014hsa}, rather than one with long-range properties, as claimed in Refs.~\cite{Sikivie:2009qn,Erken:2011dz}.) 

As seen in previous work (see Ref.~\cite{Hertzberg:2020dbk}), the merger dynamics of boson stars strongly depends on the phase of the axion field. For some phases, there is a merger, while for other phases, the boson stars do not merge. Thus, a boson star in a superposition of different phases, when coming into contact with another boson star, could initially evolve into a superposition of merged and unmerged. However, one should expect rapid decoherence during the merger event because of the stark difference in the resulting gravitational interactions of the star (merged or unmerged) with its environment.

It is common to examine boson star solutions in the nonrelativistic regime, that is the axions in orbit are moving slowly. 
From our previous analysis in Section~\ref{ParStatic}, a dilute boson star in a superposition of different density profiles, will rapidly decohere (as we already showed in Ref.~\cite{Allali:2020ttz}). However, from the analysis of Section~\ref{sec:nonrelaxion}, if it is in a superposition of different phases, then this aspect of the state will be robust against decoherence. 

On the other hand, from our analysis in Section~\ref{sec:relaxion}, if we consider dense boson stars that are made out of a bound state of (semi)-relativistic axions, then there may indeed be appreciable decoherence. Since the boson star is held together by gravity, its spatial size will not grow in time and we can ignore the spreading of the previous subsection. For the densest boson stars with negligible self-interactions, the mass and radius can be estimated as (e.g., see Ref.~\cite{Helfer:2016ljl})
\beq
M \sim {1\over G_N\,m_a},\,\,\,\,\,\,\,R\sim{1\over m_a}
\eeq
The corresponding density can be estimated as $\rho_{\DM}\sim M/R^3\sim m_a^2/G_N$. 
If we use (semi)-relativistic protons as the environment, with a realistic value of $\rho_p\sim 10^{-9}\,\mbox{GeV}/\mbox{cm}^3$, one obtains a decoherence rate of
\beq 
\Gamd \sim 10^{6}\,\mbox{s}^{-1}\!\left(\frac{\abc}{10^{-4}}\right) \left(\frac{1\,\mbox{eV}}{\ma}\right)^{\!4} \,\,\,\,\,\,\,\,\,\,(\mbox{dense boson stars})\label{GamBS}
\eeq
We note that for such dense stars the perturbation theory may breakdown and the true rate may be somewhat different, so this is an estimate. Note that we need very dense stars; if they are even moderately dilute, and their velocity dispersion is even, say, $\sim 0.1$ that of light, the exponential suppression in $K$ is huge. So dispersions that are, say, $\sim 0.5$ that of light are needed for appreciable decoherence. 
We can also consider photons from the CMB as probes; in this case, one finds a much slower decoherence rate than is caused by relativistic protons, $\Gamd \sim 10^{-5}\,\mbox{s}^{-1}\!\left(\abc/10^{-2}\right) \left(1\,\text{eV}/\ma\right)^{4}$. Thus, in the presence of (semi)-relativistic protons and CMB photons, one expects the decoherence due to the protons to dominate; so this is what sets the rate to good accuracy. 

From eq.~(\ref{GamBS}), we see that for anything other than heavy axions, the decoherence is very rapid. 
Note that this is a truly general relativistic form of quantum decoherence, since the dense boson star is described by general relativity and the environment are relativistic particles that are probing the general relativistic phenomenon of oscillations in the metric. 

\subsection{Condensates from Self-Interactions}\label{Osc}

We also mention that when there are self-interaction terms included $\sim\lambda\phi^4$, then the above solutions do not exist for $|\lambda|>G_N\,m_a^2$. Instead one is lead to other kinds of relativistic solutions. For {\em attractive} self-interactions ($\lambda<0$), these are oscillons/axitons with mass and radius given by
\beq
M \sim {m_a\over \lambda},\,\,\,\,\,\,\,R\sim{1\over m_a}
\eeq
For extremely small $\lambda$, as anticipated for an axion with $\lambda\sim m_a^2/f_a^2$ ($f_a$ is the axion decay constant, which is anticipated to be an extremely high scale, perhaps within a few orders of magnitude of the Planck scale), these masses can be appreciable too. However, in this case there is radiation in the form of scalar waves, so these states do not live too long, unless $m_a$ is extremely small. But in any case, the decoherence rate is altered from the result in eq.~(\ref{GamBS}) by a simple additional factor
\beq 
\Gamd \sim 10^{6}\,\mbox{s}^{-1}\!\left(\frac{\abc}{10^{-4}}\right) \left(\frac{1\,\mbox{eV}}{\ma}\right)^{\!4}\left(f_a\over\mpl\right)^{\!4} \,\,\,\,\,\,\,\,\,\,(\mbox{dense oscillons})
\eeq
where $\mpl\equiv1/\sqrt{G_N}$ is the Planck mass.  So unless $f_a$ is many orders of magnitude below the Planck scale, one anticipates rapid decoherence here too. Again we need very dense oscillons so that they are a bound state of semi-relativistic scalars to avoid large exponential suppression in the factor $K$. These tend to radiate fairly quickly, however.
 (In addition to the gravitational decoherence studied here, one might also consider the scalar radiation to be a form of ``environment" that one may consider tracing over; we leave this as a possible future topic.)

For completeness, we also mention that for appreciable self-interactions that are {\em repulsive} ($\lambda>0$), there are other interesting kinds of solutions that exist in which the repulsion holds the object up against gravitational collapse. However, for a real scalar, these cannot live long due to annihilation processes in its core, as determined for the first time by some of us in Ref.~\cite{Hertzberg:2020xdn}. Hence we will not explore these solutions further here. There are, however, a related set of stable solutions for a complex scalar in theories organized by a global $U(1)$ symmetry (with gravity, see Ref.~\cite{Colpi:1986ye}, and without gravity, see Ref.~\cite{Coleman:1985ki}). In this case, however, the symmetry ensures that even though the field oscillates as $\phi\propto e^{-i\omega t}$, its magnitude does not, and in turn the metric does not. Therefore even general relativity is not sensitive to the value of this phase and cannot decohere it. One may anticipate that an exact global symmetry is broken by quantum gravity, but we leave these topics for future investigation.

\subsection{Black Holes}\label{BH}

Another situation that can lead to relativistic DM would be in the region surrounding the event horizon of a black hole. In fact, for primordial black holes, one can anticipate nucleation of boson stars \cite{Hertzberg:2020hsz}. However, our analysis in this subsection does not rely on the DM forming boson stars. 

Let us imagine the following interesting sequence of events: DM organizes into some DMSCS which maintains its quantum coherence for a long time in the halo of the galaxy. Suppose it then approaches, or gets trapped by, the accretion disk around the black hole, but still far from the horizon. In this case, the accretion disk is likely to cause the spatial profile of the DMSCS to decohere using the analysis of Section \ref{ParStatic}. However, the phase can remain in a quantum superposition. Then if the DMSCS spirals in towards the horizon of the black hole it will become relativistic, and of course its environment will be relativistic here too. This means the analysis of Section \ref{sec:relaxion} becomes relevant, and one can anticpate that decoherence of the phase occurs here too. 

To estimate the decoherence rate as the DMSCS nears the horizon, we can estimate the density of accreting matter near the innermost stable circular orbit (ISCO) of a stellar mass black hole as prescribed by the Shakura-Sunyaev solutions for the accretion disk (see Ref.~\cite{Shakura:1972te}). For example, for a solar-mass black hole, at a distance $\sim 0.1\, R_S$ outside of the ISCO, where $R_S$ is the Schwarzschild radius of the black hole, the density of accreting matter is approximately $\rho_p\sim 10^{23}\,\mbox{GeV}/\mbox{cm}^3$. Further, as the DMSCS get's closer to the horizon, the density of matter increases rapidly, and the energy of the probe particles increases as they become more relativistic. Thus one expects decoherence to happen quite rapidly here. Very close to the horizon, the weak field metric analysis of this paper is not accurate, but the basic qualitative behavior indicated here may apply.

So from the point of view of coherent superpositions, it is the state that is the {\em most classical} that will ultimately cross the horizon into the black hole (although there are many other aspects to this issue, not discussed here). This may have consequences for one's thinking about the black hole information puzzle (for a review, see Ref.~\cite{Polchinski:2016hrw}) and various subtleties surrounding the quantum nature of black holes. We leave these interesting subjects for future consideration.

\subsection{Consequences for Earth Based Experiments}\label{Exp}

Let us now discuss possible implications for earth based experiments. We are thinking of haloscopes \cite{Sikivie:1983ip}, such as ADMX \cite{Du:2018uak}, which looks for direct detection of the DM axion wind that passes through the atmosphere and surface of the earth and can potentially interact with a detector through non-gravitational couplings; see next subsection for more discussion.   In our previous work \cite{Allali:2020ttz} we showed that DM passing through the earth's atmosphere leads to much quicker decoherence. This follows simply from the fact that the atmosphere of the earth is much denser that the halo of the galaxy, so the rate of particles passing through the DMSCS goes up by many orders of magnitude. This typically leads to decoherence of the spatial profile of the DMSCS, except for heavier axions. 

However, as discussed in this paper in detail, this cannot decohere the phase. Since the axion DM is expected to be nonrelativistic as it passes through the atmosphere, the rate of decoherence of the axion's phase is exponentially suppressed. This means that one should more precisely be describing an axion wave as it passes through one's experiment as
\beq
|\mbox{axion} \rangle \sim \sum_i c_i |\cos(\omega t - {\bf k}_a\cdot{\bf x}+\varphi_i)\rangle
\eeq
where we have indicated a sum over phases $\varphi_i$ (more generally, this can be lifted to an integral). This is certainly not the standard treatment that ADMX and essentially all other analyses use to make predictions for the signal. This leads to the very important question as to its possible consequences. Since these experiments are very sensitive to the axion phase, for example its phase directly impacts the phase of the electromagnetic resonance in an ADMX cavity, it can potentially be very important to note that its phase is likely in a superposition. This implies that the cavity is itself launched into a quantum superposition of different cavities with different resonant phases. On the other hand, it is highly nontrivial to see how this would itself be probed experimentally, if these ``worlds" subsequently decohere due to other interactions. This deserves further investigation.

\subsection{Non-Gravitational Interactions}\label{OtherInt}

In addition to gravitation, one can consider other interactions that couple the DM to the Standard Model particles. In the case of an axion, it enjoys an (approximate) shift symmetry and is a pseudo-scalar. So its interactions take on a specific form. At the dimension 5 level, this includes coupling to gauge bosons of the form $\Delta\mathcal{L}\propto\phi\,F_{\mu\nu}\tilde{F}^{\mu\nu}$ and coupling to fermions of the form $\Delta\mathcal{L}\propto\partial_\mu\phi\,\bar{\psi}\gamma^\mu \gamma_5\psi$. The specific couplings are model dependent \cite{Kim:1979if,Shifman:1979if,Dine:1981rt,Zhitnitsky:1980tq,Kim:2008hd,PDG2018}.

These interactions are inherently relativistic (they are particle number changing, for example) and could be studied with some of the formalism that we have outlined in this work. The resulting decoherence rate is worthy of future study. In particular, as we explained, the phase for nonrelativistic DM is robust against decoherence from gravity. It would therefore be of interest to determine its phase against these other interactions. This is especially important since earth based experiments rely upon the existence of these other interactions. So if the phase remains in a Schr\"{o}dinger cat-like state, it is important to know how it affects the experiment in question through these other operators.

\section*{Acknowledgments}
M.~P.~H. is supported in part by National Science Foundation Grant No. PHY-2013953.

\appendix

\section{Standard Field Expansion}\label{FieldExpansion}

For completeness here we mention the standard expansion of fields in Schr\"odinger picture in terms of creation and annihilation operators for a scalar
\beq 
\scal(\x)=\int \frac{ d^3 p}{(2\pi)^3}\frac{1}{\sqrt{2 E_p}}\left(\hat{a}_\pv e^{i\pv\cdot\x}+\hat{b}_\pv^\dagger e^{-i\pv\cdot\x}\right)
\eeq
\beq 
\Pi(\x)=-i\int \frac{d^3p}{(2\pi)^3}\sqrt{\frac{E_p}{2}}\left(\hat{b}_\pv e^{i\pv\cdot\x}-\hat{a}_\pv^\dagger e^{-i\pv\cdot\x}\right)
\eeq
The factors of $E_p$, defined as $E_p\equiv\sqrt{p^2+m^2}$, assure the proper relativistic normalization, so that $\scal$ transforms as a Lorentz scalar, with standard creation and annihilation operators $[\hat{a}_\pv,\hat{a}^\dagger_{\pv'}]=(2\pi)^3\,\delta^3(\pv-\pv')$.

\section{Properties of $S_{ij}$ }\label{app:sab}

The expression for $S_{ij}$ in eq.~(\ref{SijL}) seems asymmetric in the role of $\alpha$ compared to that of $\beta$, whereas one might expect some form of symmetry due to the arbitrary definitions of which coefficient is $\alpha$ and which is $\beta$. The apparent asymmetry arises from the fact that the integrals have not all been completed, and the $d^3\qt$ integral which has been completed has eliminated some $\beta$ dependence in favor of $\alpha$. To clearly see that the expected symmetry exists, let us integrate instead over $d^3q$. Then, starting from eq.~(\ref{Sijbig}) and integrating once more as in eqs.~(\ref{bint},\,\ref{rint}), the subsequent integration over $d^3q$ will replace all $q$ with a function of $\qt$ which can be defined analogously from the definition of $q_{\alpha\beta}$ as $\qt_{\beta\alpha}$. Then we are left with
\begin{align} S_{ij}=\sum_{\alpha,\beta}\int d^2\Omega\int& \frac{d^3\qt}{(2\pi)^3}\bigg\{F^*_\alpha(\qtv_\beta',\qtv_{\beta\alpha})F_\beta(\qtv'_{\beta},\qtv)\left(\frac{E_{\qt_{\beta\alpha}}+E_{\qt}+2(\alpha+\beta) \oma}{2\sqrt{E_{\qt_{\beta\alpha}} E_{\qt}}}\right)\nonumber\\
&\times e^{-i(\qtv_{\beta\alpha}-\qtv'_\beta)\cdot\Lv_i}e^{+i(\qtv-\qtv'_{\beta})\cdot\Lv_j}e^{2i(\alpha\varphi_i-\beta\varphi_j)}\psik^*(\qtv_{\beta\alpha})\psik(\qtv)\bigg\}
\end{align}
If we interchange $i$ with $j$, $\alpha$ with $\beta$, and redefine $\qt\to q$, we can write
\ba
 S_{ji}\amp=\amp\sum_{\alpha,\beta}\int d^2\Omega\int \frac{d^3q}{(2\pi)^3}\bigg\{F^*_\beta(\q_\alpha',\q_{\alpha\beta})F_\alpha(\q'_{\alpha},\q)\left(\frac{E_{q_{\alpha\beta}}+E_{q}+2(\alpha+\beta) \oma}{2\sqrt{E_{q_{\alpha\beta}} E_{q}}}\right)\nonumber\\
\amp\amp\times e^{-i(\q_{\alpha\beta}-\q'_\alpha)\cdot\Lv_j}e^{+i(\q-\q'_{\alpha})\cdot\Lv_i}e^{2i(\beta\varphi_j-\alpha\varphi_i)}\psik^*(\q_{\alpha\beta})\psik(\q)\bigg\}\nonumber\\
\amp=\amp \left(S_{ij}\right)^*\label{Sijcc}
\ea
This result should be of no surprise. The unitarity of the Schr\"{o}dinger evolution guarantees that the overlap $\braket{\psi_{s,i}^{(1)}}{\psi_{s,i}^{(1)}}$ must be real. Thus, the $S_{ii}$ must be real, which is seen by the fact that they are equal to their own complex conjugate, while the off-diagonal elements must be equal to the conjugate of their transpose.

One can continue this analysis through the analysis of Section~\ref{sec:phases}, where the sub-states of the superposition differ only in phase, and therefore the only appearance of $i$ or $j$ is in the phases. $S_{ij}$ is then broken up accordingly as in eq.~(\ref{Ssum}) and $s_{\alpha,\beta}$ is defined as in in eq.~(\ref{sab}). By inspection of eq.~(\ref{Sijcc}) and the definition of $s_{\alpha,\beta}$ in eq.~(\ref{sab}), we have
\ba 
S_{ji}\amp=\amp\sum_{\alpha,\beta}e^{2i(\beta\varphi_j-\alpha\varphi_i)}\int d^2\Omega\int \frac{d^3q}{(2\pi)^3}\,F^*_\beta(\q_\alpha',\q_{\alpha\beta})F_\alpha(\q'_{\alpha},\q)\nonumber\\
\amp\amp~~~~~~~~~~~~~~~~~~~~~~
\times\left(\frac{E_{q_{\alpha\beta}}+E_{q}+2(\alpha+\beta) \oma}{2\sqrt{E_{q_{\alpha\beta}} E_{q}}}\right)
\psik^*(\q_{\alpha\beta})\psik(\q)
\ea
Hence we have
\beq
S_{ji}=\sum_{\alpha,\beta}e^{2i(\beta\varphi_j-\alpha\varphi_i)}s_{\alpha\beta}^*=\sum_{\alpha,\beta}e^{2i(\beta\varphi_j-\alpha\varphi_i)}s_{\beta\alpha}
\eeq
and so we can conclude similarly that $s_{\alpha\beta}=s_{\beta\alpha}^*$.

\section{Generalized Cross Sections for the Gaussian Profile}\label{app:decrate estimate}

For the Gaussian profile discussed in Section~\ref{sec:gausstd} and the corresponding scattering amplitudes in eqs.~(\ref{F0gauss},\,\ref{F1gauss}), we report here on the generalized cross sections $\sig_{\alpha\beta}$.

Recall that the leading exponentials in the scattering amplitudes cause the expressions to be exponentially suppressed for large $\theta$. Thus, we will approximate the integral in eq.~(\ref{Sigmadef}) by expanding the arguments of the exponentials and the remaining parts of the integrand in powers of $\theta$ about $\theta=0$ and then taking the upper limit of $\theta$ to infinity. Then, we can obtain an expression for $\sig_{\alpha\beta}$ in terms of the parameters discussed in Section~\ref{sec:gausstd}. We quote here the result for $\sig_{1,1}$ and $\sig_{-1,-1}$(note that $\sig_{00}$ is given simply by the generalized cross section for the static metric case $\sigma_{ii}$)
\ba 
\sig_{11}\amp=\amp\frac{\pi\left(4k^2(3\mu^2-\oma^2)+2\ma^2(2k^2+m_p^2)+m_p^2(k^2+6\mu^2+2\oma E_k -k\sqrt{k^2+4\oma (\oma+E_k)})\right)^2}{32 k \mu^2 \sqrt{k^2+4\oma(\oma+E_k)}\left(k^2+2\oma(\oma+E_k)-k\sqrt{k^2+4\oma(\oma+E_k)}\right)^2}\nonumber\\
\amp\amp\times G_N^2 (\kappa M)^2 
\,e^{\frac{-2\oma\left(\oma+E_k\right)+k(-k+\sqrt{k^2+4\oma(\oma+E_k)})}{\mu^2}} 
\ea
\ba
\sig_{-1-1}\amp=\amp\frac{\pi\left(4k^2(3\mu^2-\oma^2)+2\ma^2(2k^2+m_p^2)+m_p^2(k^2+6\mu^2-2\oma E_k -k\sqrt{k^2+4\oma (\oma-E_k)})\right)^2}{32 k \mu^2 \sqrt{k^2+4\oma(\oma-E_k)}\left(k^2+2\oma(\oma-E_k)-k\sqrt{k^2+4\oma(\oma-E_k)}\right)^2}\nonumber\\
\amp\amp\times G_N^2 (\kappa M)^2 
\,e^{\frac{2\oma\left(-\oma+E_k\right)+k(-k+\sqrt{k^2+4\oma(\oma-E_k)})}{\mu^2}}
\ea
To apply these expressions to the analysis in Section~\ref{sec:relaxion}, one examines the regime in which the probe particle and the DM particle are relativistic. This is the only scenario where the exponentials in $\sig_{\alpha\beta}$ do not cause the results to be exponentially suppressed. Then, we use the fact that the energy/mass/momentum of the DM particle are assumed smaller than the energy/mass/momentum of the probe particle (this is trivially the case when, for example, the DM is a light axion and the probe is a proton). Expanding the expressions for $\sig_{\alpha\beta}$ to leading order in the DM parameters $\oma$, $\mu$, and $\ma$, we can combine the resulting expressions as in eq.~(\ref{GammaSigma}) or eq.~(\ref{GammaSigma2}) to obtain eq.~(\ref{Gammasimple}).
   
\section{Relation to Klein-Gordon Equation}\label{KGapp}

Here we explain that the one-particle Schr\"{o}dinger equation is closely related to the Klein-Gordon equation in curved spacetime, even though the latter is usually only used in the context of field theory.
For completeness, we quote here the Schr\"{o}dinger equation of Section~\ref{sec:schro}.
\beq 
(i\p_t -\sqrt{-\nabla^2+m^2})\psi(\x,t)=\left(\Phi(\x,t) \sqrt{-\nabla^2+m^2}-\frac{\Psi(\x,t) \nabla^2}{\sqrt{-\nabla^2+m^2}}\right)\psi(\x,t)\eeq

We now show that this is related to the Klein-Gordon equation in a curved background spacetime. To obtain the Klein-Gordon equation in curved spacetime, one need only generalize the flat-space Klein-Gordon equation by replacing the Minkowski metric with the curved metric $g_{\mu\nu}$ and by generalizing ordinary derivatives to covariant derivatives appropriately
\beq 
g^{\mu\nu}\nabla_\mu(\p_\nu\psi) +m^2 \psi = 0 \label{KGcurved}
\eeq

The covariant derivative in eq.~(\ref{KGcurved}) involves an ordinary derivative and a Christoffel symbol term. However, in the analysis leading to the Schr\"{o}dinger equation of Section~\ref{sec:schro}, we assumed that the metric coefficients are slowly varying in space and time. Thus, the Christoffel symbols involving spatial and time derivatives of the metric can be taken to be zero. What this means is that the variation in time of the metric is sufficiently slow compared to the variation in time of the wave function (in the case of a probe particle scattering off of an oscillating scalar field, this amounts to assuming that the energy of the probe particle is much greater than the frequency of oscillation of the scalar field), and the source is wide compared to the probe. Altogether this gives
\beq g^{\mu\nu}\p_\mu\p_\nu\psi +m^2\psi =0 \eeq
To linear order in perturbations of the metric of Section~\ref{WDM} this leads to
\beq 
\ddot{\psi}\approx(1+2\Psi+2\Phi)\nabla^2\psi -(1+2\Phi)m^2 \psi\label{KGexp}
\eeq

We can see that the solutions to the above Schr\"{o}dinger equation are compatible with the solutions of the Klein-Gordon Equation. Taking a time derivative of the Schr\"{o}dinger equation as
\beq 
-i \p_t\left(i\p_t \psi(\x,t)\right)=-i \p_t\left(H(\x,-\nabla^2)\psi(\x,t)\right)
\eeq
and then using the appropriate Hamiltonian and ignoring derivatives of the metric
\beq
\ddot{\psi}=-iH(\x,-\nabla^2)\dot{\psi}(\x,t)\approx-H^2(\x,-\nabla^2)\psi(\x,t)
\eeq
By using the previously defined Hamiltonian, we readily recover the Klein-Gordon equation (\ref{KGexp}).

Thus a time derivative of the Schr\"{o}dinger equation gives the Klein-Gordon equation. The extra differentiation promotes the Schr\"{o}dinger equation to a second-order differential equation. This introduces a new set of negative frequency solutions, which are understood to be related to the need for antiparticles for a causal interacting theory. The corresponding solutions of the Klein-Gordon equation must be restricted to the limit where only one particle's evolution is described and no particles/antiparticles are created or destroyed, and thus the negative frequency solutions (which do not solve the Schr\"{o}dinger equation) do not contribute.


\begin{thebibliography}{}

%%%%%%%%%%%%%%%%%%
%%%% Paper 1 %%%%%%%
%%%%%%%%%%%

%\cite{Allali:2020ttz}
\bibitem{Allali:2020ttz}
I.~Allali and M.~P.~Hertzberg,
``Gravitational Decoherence of Dark Matter,''
JCAP \textbf{07}, 056 (2020)
%doi:10.1088/1475-7516/2020/07/056
[arXiv:2005.12287 [gr-qc]].
%1 citations counted in INSPIRE as of 19 Nov 2020


%%%%%%%%%% DM Review. %%%%
%\cite{Peebles:2013hla}
\bibitem{Peebles:2013hla}
P.~J.~E.~Peebles,
``Dark Matter,''
Proc. Nat. Acad. Sci. \textbf{112}, 2246 (2015)
%doi:10.1073/pnas.1308786111
[arXiv:1305.6859 [astro-ph.CO]].
%21 citations counted in INSPIRE as of 23 Dec 2020


%%%%%%%%%%%%%%%%%%%%%%
%%%% Decoherence early work
%%%%%%%%%%%%%%%%%%%%%%
% Zeh1970,Zurek1981,Zurek1982

\bibitem{Zeh1970}
H. D. Zeh, 
``On the interpretation of measurement in quantum theory," 
Found. Phys. 1 69-76 (1970).

\bibitem{Zurek1981}
W. H. Zurek, 
``Pointer basis of quantum apparatus: Into what mixture does the wave packet collapse?," 
Phys. Rev. D 24 1516-1525 (1981).

\bibitem{Zurek1982}
W. H. Zurek, 
``Environment-induced superselection rules,"
Phys. Rev. D 26 1862-1880 (1982).


%%%%%%%%%%%%%%%%%%%%%%
%%%% Decoherence general formalism
%%%%%%%%%%%%%%%%%%%%%%
%Joos:1984uk,Gallis1990,Diosi1995,Giulini:1996nw,Kiefer:1997hv,Dodd:2003zk,Hornberger,Schlosshauer:2003zy,SchlosshauerBook,HornbergerIntrp,Schlosshauer:2019ewh,Nagele:2020kef

%\cite{Joos:1984uk}
\bibitem{Joos:1984uk}
E.~Joos and H.~Zeh,
``The Emergence of classical properties through interaction with the environment,''
Z. Phys. B \textbf{59}, 223-243 (1985)
%doi:10.1007/BF01725541
%425 citations counted in INSPIRE as of 15 May 2020

\bibitem{Gallis1990}
M. R. Gallis and G. N. Fleming, 
``Environmental and spontaneous localization,"
Phys. Rev. A 42, 38 (1990).

\bibitem{Diosi1995}
L. Diosi, 
``Quantum master equation of a particle in a gas environment," 
Europhys. Lett. 30 63-68 (1995).

%\cite{Giulini:1996nw}
\bibitem{Giulini:1996nw}
D.~Giulini, C.~Kiefer, E.~Joos, J.~Kupsch, I.~Stamatescu and H.~Zeh,
``Decoherence and the appearance of a classical world in quantum theory,''
2nd Edition, Springer (2003). 
%28 citations counted in INSPIRE as of 15 May 2020

%\cite{Kiefer:1997hv}
\bibitem{Kiefer:1997hv}
C.~Kiefer and E.~Joos,
``Decoherence: Concepts and examples,''
Lect. Notes Phys. \textbf{517}, 105-128 (1999)
%doi:10.1007/BFb0105342
[arXiv:quant-ph/9803052 [quant-ph]].
%14 citations counted in INSPIRE as of 15 May 2020

%\cite{Dodd:2003zk}
\bibitem{Dodd:2003zk}
P.~J.~Dodd and J.~J.~Halliwell,
``Decoherence and records for the case of a scattering environment,''
Phys. Rev. D \textbf{67}, 105018 (2003)
%doi:10.1103/PhysRevD.67.105018
[arXiv:quant-ph/0301104 [quant-ph]].
%18 citations counted in INSPIRE as of 15 May 2020

\bibitem{Hornberger}
K.~Hornberger, J.~E.~Sipe,
``Collisional decoherence reexamined,"
Phys. Rev. A 68 1094-1622 (2003) 
[arXiv:quant-ph/0303094].

%\cite{Schlosshauer:2003zy}
\bibitem{Schlosshauer:2003zy}
M.~Schlosshauer,
``Decoherence, the Measurement Problem, and Interpretations of Quantum Mechanics,''
Rev. Mod. Phys. \textbf{76}, 1267-1305 (2004)
%doi:10.1103/RevModPhys.76.1267
[arXiv:quant-ph/0312059 [quant-ph]].
%199 citations counted in INSPIRE as of 15 May 2020

\bibitem{SchlosshauerBook}
M.~A.~Schlosshauer
``Decoherence: And the Quantum-To-Classical Transition,"
ISBN 978-3-540-35773-5 Springer-Verlag (2007). 

\bibitem{HornbergerIntrp}
K.~Hornberger,
``Introduction to decoherence theory,"
Lect. Notes Phys. 768, 221-276 (2009)
[arXiv:quant-ph/0612118]. 

%\cite{Schlosshauer:2019ewh}
\bibitem{Schlosshauer:2019ewh}
M.~Schlosshauer,
``Quantum Decoherence,''
Phys. Rept. \textbf{831}, 1-57 (2019)
%doi:10.1016/j.physrep.2019.10.001
[arXiv:1911.06282 [quant-ph]].
%1 citations counted in INSPIRE as of 15 May 2020

%\cite{Nagele:2020kef}
\bibitem{Nagele:2020kef}
C.~Nagele, O.~Janssen and M.~Kleban,
``Decoherence: A Numerical Study,''
[arXiv:2010.04803 [quant-ph]].
%0 citations counted in INSPIRE as of 14 May 2021


%%%%%%%%%%%%%%%%%%%%%%
%%%% Decoherence plus gravity/cosmology below
%%%%%%%%%%%%%%%%%%%%%%
% Bassi:2017szd,Belenchia:2018szb,Asprea:2019dok,Anastopoulos:2013zya,Blencowe:2012mp,Breuer:2008rh,Shariati:2016mty,DeLisle:2019dyw,Orlando:2016pwg,Pang:2016foq,Oniga:2015lro,Bonder:2015hja,Diosi:2015vra,Colin:2014vfa,Hu:2014kia,Pikovski:2013qwa,Polarski:1995jg,Halliwell:1989vw,Kiefer:1998qe,Padmanabhan:1989rm,Kafri:2014zsa,Nelson:2016kjm,Anastopoulos:2014yja,Wang:2006vh,Kok:2003mc,Pikovski:2015wwa,Kiefer:1999gt,Mavromatos:2007hv,Tegmark:2011pi,Anastopoulos:1995ya,Colin:2014cfa,Kiefer:2008ku,Brandenberger:1990bx,Khosla:2016tss,Podolskiy:2015wna,Arrasmith:2017ogi,Albrecht:2018prr

%\cite{Bassi:2017szd}
\bibitem{Bassi:2017szd}
A.~Bassi, A.~Großardt and H.~Ulbricht,
``Gravitational Decoherence,''
Class. Quant. Grav. \textbf{34}, no.19, 193002 (2017)
%doi:10.1088/1361-6382/aa864f
[arXiv:1706.05677 [quant-ph]].
%31 citations counted in INSPIRE as of 04 May 2020

%\cite{Belenchia:2018szb}
\bibitem{Belenchia:2018szb}
A.~Belenchia, R.~M.~Wald, F.~Giacomini, E.~Castro-Ruiz, Č.~Brukner and M.~Aspelmeyer,
``Quantum Superposition of Massive Objects and the Quantization of Gravity,''
Phys. Rev. D \textbf{98}, no.12, 126009 (2018)
%doi:10.1103/PhysRevD.98.126009
[arXiv:1807.07015 [quant-ph]].
%25 citations counted in INSPIRE as of 04 May 2020

%\cite{Asprea:2019dok}
\bibitem{Asprea:2019dok}
L.~Asprea, G.~Gasbarri and A.~Bassi,
``Gravitational decoherence: a general non relativistic model,''
[arXiv:1905.01121 [quant-ph]].
%2 citations counted in INSPIRE as of 04 May 2020

%\cite{Anastopoulos:2013zya}
\bibitem{Anastopoulos:2013zya}
C.~Anastopoulos and B.~Hu,
``A Master Equation for Gravitational Decoherence: Probing the Textures of Spacetime,''
Class. Quant. Grav. \textbf{30}, 165007 (2013)
%doi:10.1088/0264-9381/30/16/165007
[arXiv:1305.5231 [gr-qc]].
%33 citations counted in INSPIRE as of 04 May 2020

%\cite{Blencowe:2012mp}
\bibitem{Blencowe:2012mp}
M.~Blencowe,
``Effective Field Theory Approach to Gravitationally Induced Decoherence,''
Phys. Rev. Lett. \textbf{111}, no.2, 021302 (2013)
%doi:10.1103/PhysRevLett.111.021302
[arXiv:1211.4751 [quant-ph]].
%47 citations counted in INSPIRE as of 04 May 2020

%\cite{Breuer:2008rh}
\bibitem{Breuer:2008rh}
H.~P.~Breuer, E.~Goklu and C.~Lammerzahl,
``Metric fluctuations and decoherence,''
Class. Quant. Grav. \textbf{26}, 105012 (2009)
%doi:10.1088/0264-9381/26/10/105012
[arXiv:0812.0420 [gr-qc]].
%20 citations counted in INSPIRE as of 04 May 2020

%\cite{Shariati:2016mty}
\bibitem{Shariati:2016mty}
A.~Shariati, M.~Khorrami and F.~Loran,
``Decoherence in quantum systems in a static gravitational field,''
EPL \textbf{115}, no.5, 50003 (2016)
%doi:10.1209/0295-5075/115/50003
[arXiv:1610.02494 [quant-ph]].
%1 citations counted in INSPIRE as of 04 May 2020

%\cite{DeLisle:2019dyw}
\bibitem{DeLisle:2019dyw}
C.~DeLisle, J.~Wilson-Gerow and P.~Stamp,
``Gravitational Decoherence, Asymptotic Quantization, and Entanglement Measures,''
[arXiv:1905.05333 [gr-qc]].
%0 citations counted in INSPIRE as of 04 May 2020

%\cite{Orlando:2016pwg}
\bibitem{Orlando:2016pwg}
P.~J.~Orlando, F.~A.~Pollock and K.~Modi,
``How does interference fall?,''
%doi:10.1007/978-3-319-53412-1_19
[arXiv:1610.02141 [quant-ph]].
%2 citations counted in INSPIRE as of 04 May 2020

%\cite{Pang:2016foq}
\bibitem{Pang:2016foq}
B.~H.~Pang, Y.~Chen and F.~Y.~Khalili,
``Universal Decoherence under Gravity: A Perspective through the Equivalence Principle,''
Phys. Rev. Lett. \textbf{117}, no.9, 090401 (2016)
%doi:10.1103/PhysRevLett.117.090401
[arXiv:1603.01984 [quant-ph]].
%16 citations counted in INSPIRE as of 04 May 2020

%\cite{Oniga:2015lro}
\bibitem{Oniga:2015lro}
T.~Oniga and C.~H.~Wang,
``Quantum gravitational decoherence of light and matter,''
Phys. Rev. D \textbf{93}, no.4, 044027 (2016)
%doi:10.1103/PhysRevD.93.044027
[arXiv:1511.06678 [quant-ph]].
%19 citations counted in INSPIRE as of 04 May 2020

%\cite{Bonder:2015hja}
\bibitem{Bonder:2015hja}
Y.~Bonder, E.~Okon and D.~Sudarsky,
``Can gravity account for the emergence of classicality?,''
Phys. Rev. D \textbf{92}, no.12, 124050 (2015)
%doi:10.1103/PhysRevD.92.124050
[arXiv:1509.04363 [gr-qc]].
%9 citations counted in INSPIRE as of 04 May 2020

%\cite{Diosi:2015vra}
\bibitem{Diosi:2015vra}
L.~Diósi,
``Centre of mass decoherence due to time dilation: paradoxical frame-dependence,''
J. Phys. Conf. Ser. \textbf{880}, no.1, 012020 (2017)
%doi:10.1088/1742-6596/880/1/012020
[arXiv:1507.05828 [quant-ph]].
%9 citations counted in INSPIRE as of 04 May 2020

%\cite{Colin:2014vfa}
\bibitem{Colin:2014vfa}
S.~Colin, T.~Durt and R.~Willox,
``Can quantum systems succumb to their own (gravitational) attraction?,''
Class. Quant. Grav. \textbf{31}, no.24, 245003 (2014)
%doi:10.1088/0264-9381/31/24/245003
[arXiv:1403.2982 [quant-ph]].
%10 citations counted in INSPIRE as of 04 May 2020

%\cite{Hu:2014kia}
\bibitem{Hu:2014kia}
B.~Hu,
``Gravitational Decoherence, Alternative Quantum Theories and Semiclassical Gravity,''
J. Phys. Conf. Ser. \textbf{504}, 012021 (2014)
%doi:10.1088/1742-6596/504/1/012021
[arXiv:1402.6584 [gr-qc]].
%16 citations counted in INSPIRE as of 04 May 2020

%\cite{Pikovski:2013qwa}
\bibitem{Pikovski:2013qwa}
I.~Pikovski, M.~Zych, F.~Costa and C.~Brukner,
``Universal decoherence due to gravitational time dilation,''
Nature Phys. \textbf{11}, 668-672 (2015)
%doi:10.1038/nphys3366
[arXiv:1311.1095 [quant-ph]].
%71 citations counted in INSPIRE as of 04 May 2020

%\cite{Polarski:1995jg}
\bibitem{Polarski:1995jg}
D.~Polarski and A.~A.~Starobinsky,
``Semiclassicality and decoherence of cosmological perturbations,''
Class. Quant. Grav. \textbf{13}, 377-392 (1996)
%doi:10.1088/0264-9381/13/3/006
[arXiv:gr-qc/9504030 [gr-qc]].
%455 citations counted in INSPIRE as of 04 May 2020

%\cite{Halliwell:1989vw}
\bibitem{Halliwell:1989vw}
J.~J.~Halliwell,
``Decoherence in Quantum Cosmology,''
Phys. Rev. D \textbf{39}, 2912 (1989)
%doi:10.1103/PhysRevD.39.2912
%205 citations counted in INSPIRE as of 04 May 2020

%\cite{Kiefer:1998qe}
\bibitem{Kiefer:1998qe}
C.~Kiefer, D.~Polarski and A.~A.~Starobinsky,
``Quantum to classical transition for fluctuations in the early universe,''
Int. J. Mod. Phys. D \textbf{7}, 455-462 (1998)
%doi:10.1142/S0218271898000292
[arXiv:gr-qc/9802003 [gr-qc]].
%157 citations counted in INSPIRE as of 04 May 2020

%\cite{Padmanabhan:1989rm}
\bibitem{Padmanabhan:1989rm}
T.~Padmanabhan,
``Decoherence in the Density Matrix Describing Quantum Three Geometries and the Emergence of Classical Space-time,''
Phys. Rev. D \textbf{39}, 2924-2932 (1989)
%doi:10.1103/PhysRevD.39.2924
%107 citations counted in INSPIRE as of 04 May 2020

%\cite{Kafri:2014zsa}
\bibitem{Kafri:2014zsa}
D.~Kafri, J.~Taylor and G.~Milburn,
``A classical channel model for gravitational decoherence,''
New J. Phys. \textbf{16}, 065020 (2014)
%doi:10.1088/1367-2630/16/6/065020
[arXiv:1401.0946 [quant-ph]].
%44 citations counted in INSPIRE as of 04 May 2020

%\cite{Nelson:2016kjm}
\bibitem{Nelson:2016kjm}
E.~Nelson,
``Quantum Decoherence During Inflation from Gravitational Nonlinearities,''
JCAP \textbf{03}, 022 (2016)
%doi:10.1088/1475-7516/2016/03/022
[arXiv:1601.03734 [gr-qc]].
%37 citations counted in INSPIRE as of 04 May 2020

%\cite{Anastopoulos:2014yja}
\bibitem{Anastopoulos:2014yja}
C.~Anastopoulos and B.~Hu,
``Problems with the Newton-Schrödinger equations,''
New J. Phys. \textbf{16}, 085007 (2014)
%doi:10.1088/1367-2630/16/8/085007
[arXiv:1403.4921 [quant-ph]].
%36 citations counted in INSPIRE as of 04 May 2020

%\cite{Wang:2006vh}
\bibitem{Wang:2006vh}
C.~H.~T.~Wang, R.~Bingham and J.~Mendonca,
``Quantum gravitational decoherence of matter waves,''
Class. Quant. Grav. \textbf{23}, L59-L65 (2006)
%doi:10.1088/0264-9381/23/18/L01
[arXiv:gr-qc/0603112 [gr-qc]].
%34 citations counted in INSPIRE as of 04 May 2020

%\cite{Kok:2003mc}
\bibitem{Kok:2003mc}
P.~Kok and U.~Yurtsever,
``Gravitational decoherence,''
Phys. Rev. D \textbf{68}, 085006 (2003)
%doi:10.1103/PhysRevD.68.085006
[arXiv:gr-qc/0306084 [gr-qc]].
%38 citations counted in INSPIRE as of 04 May 2020

%\cite{Pikovski:2015wwa}
\bibitem{Pikovski:2015wwa}
I.~Pikovski, M.~Zych, F.~Costa and C.~Brukner,
``Time dilation in quantum systems and decoherence,''
New J. Phys. \textbf{19}, no.2, 025011 (2017)
%doi:10.1088/1367-2630/aa5d92
[arXiv:1508.03296 [quant-ph]].
%26 citations counted in INSPIRE as of 04 May 2020

%\cite{Kiefer:1999gt}
\bibitem{Kiefer:1999gt}
C.~Kiefer,
``Origin of classical structure from inflation,''
Nucl. Phys. B Proc. Suppl. \textbf{88}, 255-258 (2000)
%doi:10.1016/S0920-5632(00)00779-9
[arXiv:astro-ph/0006252 [astro-ph]].
%24 citations counted in INSPIRE as of 04 May 2020

%\cite{Mavromatos:2007hv}
\bibitem{Mavromatos:2007hv}
N.~E.~Mavromatos, A.~Meregaglia, A.~Rubbia, A.~Sakharov and S.~Sarkar,
``Quantum-Gravity Decoherence Effects in Neutrino Oscillations: Expected Constraints From CNGS and J-PARC,''
Phys. Rev. D \textbf{77}, 053014 (2008)
%doi:10.1103/PhysRevD.77.053014
[arXiv:0801.0872 [hep-ph]].
%22 citations counted in INSPIRE as of 04 May 2020

%\cite{Tegmark:2011pi}
\bibitem{Tegmark:2011pi}
M.~Tegmark,
``How unitary cosmology generalizes thermodynamics and solves the inflationary entropy problem,''
Phys. Rev. D \textbf{85}, 123517 (2012)
%doi:10.1103/PhysRevD.85.123517
[arXiv:1108.3080 [hep-th]].
%20 citations counted in INSPIRE as of 04 May 2020

%\cite{Anastopoulos:1995ya}
\bibitem{Anastopoulos:1995ya}
C.~Anastopoulos,
``Quantum theory of nonrelativistic particles interacting with gravity,''
Phys. Rev. D \textbf{54}, 1600-1605 (1996)
%doi:10.1103/PhysRevD.54.1600
[arXiv:gr-qc/9511004 [gr-qc]].
%16 citations counted in INSPIRE as of 04 May 2020

%\cite{Colin:2014cfa}
\bibitem{Colin:2014cfa}
S.~Colin, T.~Durt and R.~Willox,
``Crucial tests of macrorealist and semiclassical gravity models with freely falling mesoscopic nanospheres,''
Phys. Rev. A \textbf{93}, no.6, 062102 (2016)
%doi:10.1103/PhysRevA.93.062102
[arXiv:1402.5653 [quant-ph]].
%13 citations counted in INSPIRE as of 04 May 2020

%\cite{Kiefer:2008ku}
\bibitem{Kiefer:2008ku}
C.~Kiefer and D.~Polarski,
``Why do cosmological perturbations look classical to us?,''
Adv. Sci. Lett. \textbf{2}, 164-173 (2009)
%doi:10.1166/asl.2009.1023
[arXiv:0810.0087 [astro-ph]].
%124 citations counted in INSPIRE as of 15 May 2020

%\cite{Brandenberger:1990bx}
\bibitem{Brandenberger:1990bx}
R.~H.~Brandenberger, R.~Laflamme and M.~Mijic,
``Classical Perturbations From Decoherence of Quantum Fluctuations in the Inflationary Universe,''
Mod. Phys. Lett. A \textbf{5}, 2311-2318 (1990)
%doi:10.1142/S0217732390002651
%55 citations counted in INSPIRE as of 15 May 2020

%\cite{Khosla:2016tss}
\bibitem{Khosla:2016tss}
K.~Khosla and N.~Altamirano,
``Detecting gravitational decoherence with clocks: Limits on temporal resolution from a classical channel model of gravity,''
Phys. Rev. A \textbf{95}, no.5, 052116 (2017)
%doi:10.1103/PhysRevA.95.052116
[arXiv:1611.09919 [quant-ph]].
%13 citations counted in INSPIRE as of 08 Jan 2021

%\cite{Podolskiy:2015wna}
\bibitem{Podolskiy:2015wna}
D.~Podolskiy and R.~Lanza,
``On decoherence in quantum gravity,''
Annalen Phys. \textbf{528}, no.9-10, 663-676 (2016)
%doi:10.1002/andp.201600011
[arXiv:1508.05377 [gr-qc]].
%1 citations counted in INSPIRE as of 08 Jan 2021

%\cite{Arrasmith:2017ogi}
\bibitem{Arrasmith:2017ogi}
A.~Arrasmith, A.~Albrecht and W.~H.~Zurek,
``Decoherence of black hole superpositions by Hawking radiation,''
Nature Commun. \textbf{10}, no.1, 1024 (2019)
%doi:10.1038/s41467-019-08426-4
[arXiv:1708.09353 [quant-ph]].
%4 citations counted in INSPIRE as of 08 Jan 2021

%\cite{Albrecht:2018prr}
\bibitem{Albrecht:2018prr}
A.~Albrecht, S.~Kanno and M.~Sasaki,
``Quantum entanglement in de Sitter space with a wall, and the decoherence of bubble universes,''
Phys. Rev. D \textbf{97}, no.8, 083520 (2018)
%doi:10.1103/PhysRevD.97.083520
[arXiv:1802.08794 [hep-th]].
%11 citations counted in INSPIRE as of 08 Jan 2021


%%%%%%%%%%%%%%%%%%%%%%
%%%%% chaos
%%%%%%%%%%%%%%%%%%%%%%
% Albrecht:2012zp

%\cite{Albrecht:2012zp}
\bibitem{Albrecht:2012zp}
A.~Albrecht and D.~Phillips,
``Origin of probabilities and their application to the multiverse,''
Phys. Rev. D \textbf{90}, no.12, 123514 (2014)
%doi:10.1103/PhysRevD.90.123514
[arXiv:1212.0953 [gr-qc]].
%8 citations counted in INSPIRE as of 17 May 2020



%%%%%%%%%%%%%%%%%%%%%%
%%%%%%%% Axions Basics
%%%%%%%%%%%%%%%%%%%%%%%
% Peccei:1977hh,Weinberg:1977ma,Wilczek:1977pj
% Preskill:1982cy,Abbott:1982af,Dine:1982ah,AxionBook,Jaeckel:2010ni

%\cite{Peccei:1977hh}
\bibitem{Peccei:1977hh}
R.~Peccei and H.~R.~Quinn,
``CP Conservation in the Presence of Instantons,''
Phys. Rev. Lett. \textbf{38}, 1440-1443 (1977)
%doi:10.1103/PhysRevLett.38.1440
%5265 citations counted in INSPIRE as of 16 May 2020

%\cite{Weinberg:1977ma}
\bibitem{Weinberg:1977ma}
S.~Weinberg,
``A New Light Boson?,''
Phys. Rev. Lett. \textbf{40}, 223-226 (1978)
%doi:10.1103/PhysRevLett.40.223
%3653 citations counted in INSPIRE as of 16 May 2020

%\cite{Wilczek:1977pj}
\bibitem{Wilczek:1977pj}
F.~Wilczek,
``Problem of Strong  $P$  and  $T$  Invariance in the Presence of Instantons,''
Phys. Rev. Lett. \textbf{40}, 279-282 (1978)
%doi:10.1103/PhysRevLett.40.279
%3510 citations counted in INSPIRE as of 16 May 2020

%\cite{Preskill:1982cy}
\bibitem{Preskill:1982cy}
J.~Preskill, M.~B.~Wise and F.~Wilczek,
``Cosmology of the Invisible Axion,''
Phys. Lett. B \textbf{120}, 127-132 (1983)
%doi:10.1016/0370-2693(83)90637-8
%1910 citations counted in INSPIRE as of 16 May 2020

%\cite{Abbott:1982af}
\bibitem{Abbott:1982af}
L.~Abbott and P.~Sikivie,
``A Cosmological Bound on the Invisible Axion,''
Phys. Lett. B \textbf{120}, 133-136 (1983)
%doi:10.1016/0370-2693(83)90638-X
%1792 citations counted in INSPIRE as of 16 May 2020

%\cite{Dine:1982ah}
\bibitem{Dine:1982ah}
M.~Dine and W.~Fischler,
``The Not So Harmless Axion,''
Phys. Lett. B \textbf{120}, 137-141 (1983)
%doi:10.1016/0370-2693(83)90639-1
%1757 citations counted in INSPIRE as of 16 May 2020

\bibitem{AxionBook}
M. Kuster, G. Raffelt, B. Beltran,
``Axions: Theory, Cosmology, and Experimental Searches,"
Springer, (2007).

%\cite{Jaeckel:2010ni}
\bibitem{Jaeckel:2010ni}
J.~Jaeckel and A.~Ringwald,
``The Low-Energy Frontier of Particle Physics,''
Ann. Rev. Nucl. Part. Sci. \textbf{60}, 405-437 (2010)
%doi:10.1146/annurev.nucl.012809.104433
[arXiv:1002.0329 [hep-ph]].
%652 citations counted in INSPIRE as of 18 May 2020


%%%%% axion classical vs quantum %%%%%%
% Davidson:2014hfa,Guth:2014hsa
%Hertzberg:2016tal

%\cite{Davidson:2014hfa}
\bibitem{Davidson:2014hfa}
S.~Davidson,
``Axions: Bose Einstein Condensate or Classical Field?,''
Astropart. Phys. \textbf{65}, 101-107 (2015)
%doi:10.1016/j.astropartphys.2014.12.007
[arXiv:1405.1139 [hep-ph]].
%47 citations counted in INSPIRE as of 18 May 2020

%\cite{Guth:2014hsa}
\bibitem{Guth:2014hsa}
A.~H.~Guth, M.~P.~Hertzberg and C.~Prescod-Weinstein,
``Do Dark Matter Axions Form a Condensate with Long-Range Correlation?,''
Phys. Rev. D \textbf{92}, no.10, 103513 (2015)
%doi:10.1103/PhysRevD.92.103513
[arXiv:1412.5930 [astro-ph.CO]].
%135 citations counted in INSPIRE as of 18 May 2020

%\cite{Hertzberg:2016tal}
\bibitem{Hertzberg:2016tal}
M.~P.~Hertzberg,
``Quantum and Classical Behavior in Interacting Bosonic Systems,''
JCAP \textbf{11}, 037 (2016)
%doi:10.1088/1475-7516/2016/11/037
[arXiv:1609.01342 [hep-ph]].
%22 citations counted in INSPIRE as of 17 May 2020




%%%%%%%%%%%%%%%%%%%%%%
%%%%% Scattering theory
%%%%%%%%%%%%%%%%%%%%%%
% Sakurai,MurayamaWebpage,Norsen,Ishikawa,Karlovets:2015nva

\bibitem{Sakurai}
J. J. Sakurai,
``Modern Quantum Mechanics (Revised Edition)"
ISBN-10: 0201539292, Addison Wesley (1993). 

\bibitem{MurayamaWebpage}
H. Murayama, 
Quantum Mechanics II notes, 
http://hitoshi.berkeley.edu/221B/index.html

%\bibitem{Norsen}
%T.~Norsen, J.~Lande, S. B. McKagan,
%``How and why to think about scattering in terms of wave packets instead of plane waves,"
%arXiv:0808.3566 [quant-ph] (2008). 

%\bibitem{Ishikawa}
%K.~Ishikawa, Y.~Tobita,
%``On coherence lengths of wave packets,"
%Prog. Theor. Phys. 122 (2009), 1111-1136
%[arXiv:0906.3938 [quant-ph]].

%\cite{Karlovets:2015nva}
%\bibitem{Karlovets:2015nva}
%D.~Karlovets, G.~Kotkin and V.~Serbo,
%``Born approximation for scattering of wave packets on atoms. I. 
%Theoretical background for scattering of a wave packet on a potential field,''
%Phys. Rev. A \textbf{92}, 052703 (2015)
%doi:10.1103/PhysRevA.92.052703
%[arXiv:1508.00026 [quant-ph]].
%8 citations counted in INSPIRE as of 15 May 2020

% galactic velocities
%
%\cite{Kuhlen:2012ft}
\bibitem{Kuhlen:2012ft}
M.~Kuhlen, M.~Vogelsberger and R.~Angulo,
``Numerical Simulations of the Dark Universe: State of the Art and the Next Decade,''
Phys. Dark Univ. \textbf{1}, 50-93 (2012)
%doi:10.1016/j.dark.2012.10.002
[arXiv:1209.5745 [astro-ph.CO]].
%92 citations counted in INSPIRE as of 17 May 2020

%%%%% Time dependent 1+1 dimension
\bibitem{Byrd2012}
T.~A.~Byrd, M.~K.~Ivory, A.~J.~Pyle, S.~Aubin, K.~A.~Mitchell, J.~B.~Delos and K.~K.~Das
``Scattering by an oscillating barrier: quantum, classical, and semiclassical comparison,"
Phys.~Rev.~ A, {\bf 86}, 013622 (2012)
%DOI: 10.1103/PhysRevA.86.013622
[arXiv:1205.1484 [cond-mat.quant-gas]].



%%%%%%%%%%%%%%%%%%%%%
%%%%%%%% Galactic properties
%%%%%%%%%%%%%%%%%%%%%

%\cite{Read:2014qva}
\bibitem{Read:2014qva}
J.~Read,
``The Local Dark Matter Density,''
J. Phys. G \textbf{41}, 063101 (2014)
%doi:10.1088/0954-3899/41/6/063101
[arXiv:1404.1938 [astro-ph.GA]].
%257 citations counted in INSPIRE as of 16 May 2020


% density of cosmic rays
%
%\cite{Persic:2017qxo}
\bibitem{Persic:2017qxo}
M.~Persic and Y.~Rephaeli,
``Cosmic-ray energy densities in star-forming galaxies,''
EPJ Web Conf. \textbf{136}, 02008 (2017)
%doi:10.1051/epjconf/201713602008
%0 citations counted in INSPIRE as of 22 Dec 2020


%%%%Gaussian spreading in KG%%%%%%%

%\cite{Naumov:2013uia}
\bibitem{Naumov:2013uia}
D.~V.~Naumov,
``On the Theory of Wave Packets,''
Phys. Part. Nucl. Lett. \textbf{10}, 642-650 (2013)
%doi:10.1134/S1547477113070145
[arXiv:1309.1717 [quant-ph]].
%24 citations counted in INSPIRE as of 15 Dec 2020



%%%%%%%%%%%%%%%%%%%%%%
%%%%%%%%Axion stars
%%%%%%%%%%%%%%%%%%%%%%
% Tkachev:1986tr,Gleiser:1988rq,Seidel:1990jh,Tkachev:1991ka,Jetzer:1991jr,Kolb:1993zz,Liddle:1993ha,Sharma:2008sc,Chavanis:2011zi,Chavanis:2011zm,Liebling:2012fv,Visinelli:2017ooc,Schiappacasse:2017ham,Hertzberg:2018lmt,Hertzberg:2018zte,Levkov:2018kau,Hertzberg:2020dbk

%\cite{Tkachev:1986tr}
\bibitem{Tkachev:1986tr}
I.~Tkachev,
``Coherent scalar field oscillations forming compact astrophysical objects,''
Sov. Astron. Lett. \textbf{12}, 305-308 (1986)
%71 citations counted in INSPIRE as of 19 May 2020

%\cite{Gleiser:1988rq}
\bibitem{Gleiser:1988rq}
M.~Gleiser,
``Stability of Boson Stars,''
Phys. Rev. D \textbf{38}, 2376 (1988)
%doi:10.1103/PhysRevD.38.2376
%139 citations counted in INSPIRE as of 19 May 2020

%\cite{Seidel:1990jh}
\bibitem{Seidel:1990jh}
E.~Seidel and W.~M.~Suen,
``Dynamical Evolution of Boson Stars. 1. Perturbing the Ground State,''
Phys. Rev. D \textbf{42}, 384-403 (1990)
%doi:10.1103/PhysRevD.42.384
%186 citations counted in INSPIRE as of 19 May 2020

%\cite{Tkachev:1991ka}
\bibitem{Tkachev:1991ka}
I.~Tkachev,
``On the possibility of Bose star formation,''
Phys. Lett. B \textbf{261}, 289-293 (1991)
%doi:10.1016/0370-2693(91)90330-S
%73 citations counted in INSPIRE as of 19 May 2020

%\cite{Jetzer:1991jr}
\bibitem{Jetzer:1991jr}
P.~Jetzer,
``Boson stars,''
Phys. Rept. \textbf{220}, 163-227 (1992)
%doi:10.1016/0370-1573(92)90123-H
%292 citations counted in INSPIRE as of 19 May 2020

%\cite{Liddle:1993ha}
\bibitem{Liddle:1993ha}
A.~R.~Liddle and M.~S.~Madsen,
``The Structure and formation of boson stars,''
Int. J. Mod. Phys. D \textbf{1}, 101-144 (1992)
%doi:10.1142/S0218271892000057
%127 citations counted in INSPIRE as of 19 May 2020

%\cite{Kolb:1993zz}
\bibitem{Kolb:1993zz}
E.~W.~Kolb and I.~I.~Tkachev,
``Axion miniclusters and Bose stars,''
Phys. Rev. Lett. \textbf{71}, 3051-3054 (1993)
%doi:10.1103/PhysRevLett.71.3051
[arXiv:hep-ph/9303313 [hep-ph]].
%216 citations counted in INSPIRE as of 19 May 2020

%\cite{Sharma:2008sc}
\bibitem{Sharma:2008sc}
R.~Sharma, S.~Karmakar and S.~Mukherjee,
``Boson star and dark matter,''
[arXiv:0812.3470 [gr-qc]].
%17 citations counted in INSPIRE as of 19 May 2020

%\cite{Chavanis:2011zi}
\bibitem{Chavanis:2011zi}
P.~H.~Chavanis,
``Mass-radius relation of Newtonian self-gravitating Bose-Einstein condensates with short-range interactions: I. Analytical results,''
Phys. Rev. D \textbf{84}, 043531 (2011)
%doi:10.1103/PhysRevD.84.043531
[arXiv:1103.2050 [astro-ph.CO]].
%199 citations counted in INSPIRE as of 19 May 2020

%\cite{Chavanis:2011zm}
\bibitem{Chavanis:2011zm}
P.~Chavanis and L.~Delfini,
``Mass-radius relation of Newtonian self-gravitating Bose-Einstein condensates with short-range interactions: II. Numerical results,''
Phys. Rev. D \textbf{84}, 043532 (2011)
%doi:10.1103/PhysRevD.84.043532
[arXiv:1103.2054 [astro-ph.CO]].
%137 citations counted in INSPIRE as of 19 May 2020

%\cite{Liebling:2012fv}
\bibitem{Liebling:2012fv}
S.~L.~Liebling and C.~Palenzuela,
``Dynamical Boson Stars,''
Living Rev. Rel. \textbf{20}, no.1, 5 (2017)
%doi:10.12942/lrr-2012-6
[arXiv:1202.5809 [gr-qc]].
%274 citations counted in INSPIRE as of 19 May 2020

%\cite{Schiappacasse:2017ham}
\bibitem{Schiappacasse:2017ham}
E.~D.~Schiappacasse and M.~P.~Hertzberg,
``Analysis of Dark Matter Axion Clumps with Spherical Symmetry,''
JCAP \textbf{01}, 037 (2018)
%doi:10.1088/1475-7516/2018/01/037
[arXiv:1710.04729 [hep-ph]].
%43 citations counted in INSPIRE as of 19 May 2020

%\cite{Visinelli:2017ooc}
\bibitem{Visinelli:2017ooc}
L.~Visinelli, S.~Baum, J.~Redondo, K.~Freese and F.~Wilczek,
``Dilute and dense axion stars,''
Phys. Lett. B \textbf{777}, 64-72 (2018)
%doi:10.1016/j.physletb.2017.12.010
[arXiv:1710.08910 [astro-ph.CO]].
%74 citations counted in INSPIRE as of 19 May 2020

%\cite{Hertzberg:2018lmt}
\bibitem{Hertzberg:2018lmt}
M.~P.~Hertzberg and E.~D.~Schiappacasse,
``Scalar dark matter clumps with angular momentum,''
JCAP \textbf{08}, 028 (2018)
%doi:10.1088/1475-7516/2018/08/028
[arXiv:1804.07255 [hep-ph]].
%16 citations counted in INSPIRE as of 19 May 2020

%\cite{Levkov:2018kau}
\bibitem{Levkov:2018kau}
D.~Levkov, A.~Panin and I.~Tkachev,
``Gravitational Bose-Einstein condensation in the kinetic regime,''
Phys. Rev. Lett. \textbf{121}, no.15, 151301 (2018)
%doi:10.1103/PhysRevLett.121.151301
[arXiv:1804.05857 [astro-ph.CO]].
%52 citations counted in INSPIRE as of 16 May 2020

%\cite{Hertzberg:2018zte}
\bibitem{Hertzberg:2018zte}
M.~P.~Hertzberg and E.~D.~Schiappacasse,
``Dark Matter Axion Clump Resonance of Photons,''
JCAP \textbf{11}, 004 (2018)
%doi:10.1088/1475-7516/2018/11/004
[arXiv:1805.00430 [hep-ph]].
%32 citations counted in INSPIRE as of 19 May 2020

%\cite{Hertzberg:2020dbk}
\bibitem{Hertzberg:2020dbk}
M.~P.~Hertzberg, Y.~Li and E.~D.~Schiappacasse,
``Merger of Dark Matter Axion Clumps and Resonant Photon Emission,''
JCAP \textbf{07}, 067 (2020)
%doi:10.1088/1475-7516/2020/07/067
[arXiv:2005.02405 [hep-ph]].
%20 citations counted in INSPIRE as of 24 May 2021

%%%%%%%%%%%%%%%%%%%%%
%%%%%% quantum BEC %%%%%
%%%%%%%%%%%%%%%%%%%
% Sikivie:2009qn,Erken:2011dz

%\cite{Sikivie:2009qn}
\bibitem{Sikivie:2009qn}
P.~Sikivie and Q.~Yang,
``Bose-Einstein Condensation of Dark Matter Axions,''
Phys. Rev. Lett. \textbf{103}, 111301 (2009)
%doi:10.1103/PhysRevLett.103.111301
[arXiv:0901.1106 [hep-ph]].
%304 citations counted in INSPIRE as of 16 May 2020

%\cite{Erken:2011dz}
\bibitem{Erken:2011dz}
O.~Erken, P.~Sikivie, H.~Tam and Q.~Yang,
``Cosmic axion thermalization,''
Phys. Rev. D \textbf{85}, 063520 (2012)
%doi:10.1103/PhysRevD.85.063520
[arXiv:1111.1157 [astro-ph.CO]].
%107 citations counted in INSPIRE as of 16 May 2020


% maximum mass boson star
%
% %\cite{Helfer:2016ljl}
\bibitem{Helfer:2016ljl}
T.~Helfer, D.~J.~E.~Marsh, K.~Clough, M.~Fairbairn, E.~A.~Lim and R.~Becerril,
``Black hole formation from axion stars,''
JCAP \textbf{03}, 055 (2017)
%doi:10.1088/1475-7516/2017/03/055
[arXiv:1609.04724 [astro-ph.CO]].
%86 citations counted in INSPIRE as of 22 Dec 2020

% repulsive interactions
%
%\cite{Hertzberg:2020xdn}
\bibitem{Hertzberg:2020xdn}
M.~P.~Hertzberg, F.~Rompineve and J.~Yang,
``Decay of Boson Stars with Application to Glueballs and Other Real Scalars,''
Phys. Rev. D \textbf{103}, no.2, 023536 (2021)
%doi:10.1103/PhysRevD.103.023536
[arXiv:2010.07927 [hep-ph]].
%7 citations counted in INSPIRE as of 24 May 2021

% complex scalars
%
%\cite{Colpi:1986ye}
\bibitem{Colpi:1986ye}
M.~Colpi, S.~L.~Shapiro and I.~Wasserman,
``Boson Stars: Gravitational Equilibria of Selfinteracting Scalar Fields,''
Phys. Rev. Lett. \textbf{57}, 2485-2488 (1986)
%doi:10.1103/PhysRevLett.57.2485
%446 citations counted in INSPIRE as of 22 Dec 2020
%
%\cite{Coleman:1985ki}
\bibitem{Coleman:1985ki}
S.~R.~Coleman,
``Q Balls,''
Nucl. Phys. B \textbf{262}, 263 (1985)
[erratum: Nucl. Phys. B \textbf{269}, 744 (1986)]
%doi:10.1016/0550-3213(86)90520-1
%825 citations counted in INSPIRE as of 22 Dec 2020


% near black holes
%
%\cite{Hertzberg:2020hsz}
\bibitem{Hertzberg:2020hsz}
M.~P.~Hertzberg, E.~D.~Schiappacasse and T.~T.~Yanagida,
``Axion Star Nucleation in Dark Minihalos around Primordial Black Holes,''
Phys. Rev. D \textbf{102}, no.2, 023013 (2020)
%doi:10.1103/PhysRevD.102.023013
[arXiv:2001.07476 [astro-ph.CO]].
%6 citations counted in INSPIRE as of 22 Dec 2020

%Accretion Disk%%%%%%%%%%%%%%%%%%%
%
%\cite{Shakura:1972te}
\bibitem{Shakura:1972te}
N.~I.~Shakura and R.~A.~Sunyaev,
``Black holes in binary systems. Observational appearance,''
Astron. Astrophys. \textbf{24}, 337-355 (1973)
%4745 citations counted in INSPIRE as of 22 Dec 2020

% Information paradox
%
%\cite{Polchinski:2016hrw}
\bibitem{Polchinski:2016hrw}
J.~Polchinski,
``The Black Hole Information Problem,''
%doi:10.1142/9789813149441\_0006
[arXiv:1609.04036 [hep-th]].
%82 citations counted in INSPIRE as of 22 Dec 2020


% haloscopes and ADMX
%
%\cite{Sikivie:1983ip}
\bibitem{Sikivie:1983ip}
P.~Sikivie,
``Experimental Tests of the Invisible Axion,''
Phys. Rev. Lett. \textbf{51}, 1415-1417 (1983)
[erratum: Phys. Rev. Lett. \textbf{52}, 695 (1984)]
%doi:10.1103/PhysRevLett.51.1415
%1455 citations counted in INSPIRE as of 22 Dec 2020
%
%\cite{Du:2018uak}
\bibitem{Du:2018uak}
N.~Du \textit{et al.} [ADMX],
``A Search for Invisible Axion Dark Matter with the Axion Dark Matter Experiment,''
Phys. Rev. Lett. \textbf{120}, no.15, 151301 (2018)
%doi:10.1103/PhysRevLett.120.151301
[arXiv:1804.05750 [hep-ex]].
%245 citations counted in INSPIRE as of 22 Dec 2020

%%%%%%%%%%%%%%%%%%%%%%
%%%%%%%% Axions Models
%%%%%%%%%%%%%%%%%%%%%%%
% Kim:1979if,Shifman:1979if,Dine:1981rt,Zhitnitsky:1980tq,Kim:2008hd,PDG2018


%\cite{Kim:1979if}
\bibitem{Kim:1979if}
J.~E.~Kim,
``Weak Interaction Singlet and Strong CP Invariance,''
Phys. Rev. Lett. \textbf{43}, 103 (1979)
%doi:10.1103/PhysRevLett.43.103
%2127 citations counted in INSPIRE as of 16 May 2020

%\cite{Shifman:1979if}
\bibitem{Shifman:1979if}
M.~A.~Shifman, A.~Vainshtein and V.~I.~Zakharov,
``Can Confinement Ensure Natural CP Invariance of Strong Interactions?,''
Nucl. Phys. B \textbf{166}, 493-506 (1980)
%doi:10.1016/0550-3213(80)90209-6
%1842 citations counted in INSPIRE as of 16 May 2020

%\cite{Dine:1981rt}
\bibitem{Dine:1981rt}
M.~Dine, W.~Fischler and M.~Srednicki,
``A Simple Solution to the Strong CP Problem with a Harmless Axion,''
Phys. Lett. B \textbf{104}, 199-202 (1981)
%doi:10.1016/0370-2693(81)90590-6
%2546 citations counted in INSPIRE as of 16 May 2020

%\cite{Zhitnitsky:1980tq}
\bibitem{Zhitnitsky:1980tq}
A.~Zhitnitsky,
``On Possible Suppression of the Axion Hadron Interactions. (In Russian),''
Sov. J. Nucl. Phys. \textbf{31}, 260 (1980)
%1468 citations counted in INSPIRE as of 16 May 2020

%\cite{Kim:2008hd}
\bibitem{Kim:2008hd}
J.~E.~Kim and G.~Carosi,
``Axions and the Strong CP Problem,''
Rev. Mod. Phys. \textbf{82}, 557-602 (2010)
%doi:10.1103/RevModPhys.82.557
[arXiv:0807.3125 [hep-ph]].
%639 citations counted in INSPIRE as of 16 May 2020

%%%%%%%% review
\bibitem{PDG2018}
M. Tanabashi et al. (Particle Data Group), 
Phys. Rev. D 98, 030001 (2018) and 2019 update.


\end{thebibliography}
\end{document}